\documentclass[pre,10pt,twocolumn,superscriptaddress,floatfix,noeprint]{revtex4-1}

\usepackage{amsmath}
\usepackage{latexsym}
\usepackage{amssymb}
\usepackage{graphics,epstopdf}
\usepackage{hyperref}
\usepackage[caption=false]{subfig}
\usepackage[percent]{overpic}

\hypersetup{
	colorlinks = true,
	linkcolor =blue,
	citecolor=blue, 
	urlcolor=blue 
}

\usepackage{mathtools}

\newcommand{\sandwich}[3]{\langle #1 \vert #2 \vert #3 \rangle}

\newcommand{\Tr}{\ensuremath{\mathrm{Tr}\,}}
\newcommand{\avg}[1]{\ensuremath{\langle #1 \rangle}}

\newcommand{\ket}[1]{\ensuremath{\vert #1 \rangle}}
\newcommand{\bra}[1]{\ensuremath{\langle #1 \vert}}

\MakeRobust{\eqref}

\newcommand{\hs}{\hat{\sigma}}

\newcommand{\hrho}{\hat{\rho}}
\newcommand{\Hop}{\hat{H}}
\newcommand{\hO}{\hat{O}}
\newcommand{\mD}{\mathcal{D}}
\newcommand{\mL}{\mathcal{L}}
\newcommand{\Uop}{\hat{U}}

\newcommand{\aop}{\hat{a}}
\newcommand{\adop}{\hat{a}^{\dagger}}
\newcommand{\pop}{\hat{p}}
\newcommand{\xop}{\hat{x}}

\newcommand{\sz}{\hat{\sigma}_z}
\newcommand{\sx}{\hat{\sigma}_x}

\newcommand{\Piop}{\hat{\Pi}}

\DeclareMathOperator{\cosech}{cosech}

\let\oldcdot\cdot
\usepackage{breqn}
\let\cdot\oldcdot

\makeatletter
\let\cat@comma@active\@empty
\makeatother
\usepackage{marginnote}
\begin{document}
\title{Optimization of Asymmetric Quantum Otto Engine Cycles}
\author{Rahul Shastri}
\email{shastri.rahul@iitgn.ac.in}
\affiliation{Indian Institute of Technology Gandhinagar, Palaj, Gujarat 382355, India}
\author{B. Prasanna Venkatesh}
\email{prasanna.b@iitgn.ac.in}
\affiliation{Indian Institute of Technology Gandhinagar, Palaj, Gujarat 382355, India}
\begin{abstract}
We consider the optimization of the work output and fluctuations of a finite-time quantum Otto heat engine cycle consisting of compression and expansion work strokes of unequal duration. The asymmetry of the cycle is characterized by a parameter $r_u$ giving the ratio of the times for the compression and expansion work strokes. For such an asymmetric quantum Otto engine cycle, with working substance chosen as a harmonic oscillator or a two-level system, we find that the optimal values of $r_u$ maximising the work output and the reliability (defined as the ratio of average work output to its standard deviation) shows discontinuities as a function of the total time taken for the cycle. Moreover we identify cycles of some specific duration where both the work output and the reliability take their largest values for the same value of the asymmetry parameter $r_u$.
\end{abstract}

\maketitle


\section{\label{sec:Introduction}Introduction}

Heat engines have served as the primary technological applications as well as useful test-beds for the illustration of the laws of classical thermodynamics. This feature has also extended to the field of quantum thermodynamics with several pioneering contributions \cite{Scovil1959,Alicki1979} based on the study of heat engines with a quantum system as the working substance. A significant amount of effort has gone into identifying set-ups and scenarios where such quantum heat engines with a variety of working substance systems and operated according to different thermodynamic cycles provide performance advantages in terms of \emph{average} work output and power \cite{Vinjanampathy2016,Binder2018,Bhattacharjee2021,Mukherjee2021,Myers2022}. More recently, the community has turned its attention also towards the study of fluctuations of the central figures of merit such as work output and efficiency of such engines \cite{Denzler20,Saryal21,Saryal21a}. Here, the discovery of the far-reaching Thermodynamic Uncertainty Relations (TURs) bounding the relation between fluctuations and entropy production \cite{Barato15,Gingrich16,Timpanaro19} have both been an inspiration and guidance for several results considering specific thermal machine cycles and set ups \cite{Pietzonka18,Saryal21c,Saryal21a,Saryal21}. Moreover, in a parallel development there is also a concerted attempt to find configurations and protocols that optimize different performance metrics for quantum heat engines and thermal machines. A non-exhaustive list of advances include designing protocols that maximise efficiency and power for quantum Stirling engine cycle \cite{Feldmann2000}, finite time quantum Carnot engine cycle \cite{Allahverdyan2013,Cavina2017,Abiuso2020,Dann2020}, and finite time quantum Otto engine cycle \cite{Rezek2006,Quan2007,Abah2012,Karimi2016,Kosloff2017,Chen2019,Das2020,Insinga2020,Singh2020,Zhang2020}. Some recent advances include attempts to design generalized Otto cycles based on fast work strokes \cite{Cavina2021}, using geometric approaches to optimize adiabatic quantum machines \cite{TerrenAlonso2022}, and even employing machine-learning based techniques to find optimal configurations for heat engines \cite{Ashida2021,Erdman2021,Khait2021}. 

In this article, we present an optimization approach for a particular quantum heat engine cycle. We consider a finite-time quantum Otto heat engine cycle with compression and expansion work strokes as well as hot and cold thermal bath isochores of unequal duration. Such a set up, introduced in \cite{PhysRevE.94.012137}, is different from standard Otto cycles with equal time compression and expansion strokes. In \cite{PhysRevE.94.012137}, for a harmonic oscillator (HO) working substance, the ratio of the duration between the compression and expansion work strokes characterizing the \emph{asymmetry} between the strokes was optimized to give the maximum work output and efficiency. It was found that the optimal value of this asymmetry parameter (to be defined more precisely below) shows discontinuities as a function of the total time for the engine cycle. Here, we consider the optimization of the average work $\avg{w}$ as well as a measure of the fluctuations in the work output called reliability $R_w$ which is defined as the ratio of the average of work and its standard deviation. We find the following central results. The optimal value of the asymmetry parameter maximising reliability $R_w$ shows discontinuities much like the one maximising the average work output. Secondly, while the value of the asymmetry parameter for which the work output is maximised is not same as the one where reliability is maximised we find that for some specific choices of the total time of the cycle the optimal values coincide. Finally, we find that the discontinuities as well as the \emph{co-optimization} of average work output and reliability are also present for the cycle with a two-level system (TLS) as the working substance hinting at the generality of the results.

This article is structured as follows. In Sec.~\eqref{sec:Model} we describe the finite-time asymmetric quantum Otto engine (a-QOE) setup for a general working substance system. We provide formal expressions for the work and heat statistics in this scenario and for the key quantities that we seek to optimize - the average work output $\avg{w}$ and the reliability $R_w$. In Sec.~\eqref{sec:System}, we focus on two specific realizations of the a-QOE with a harmonic oscillator and two-level system as working substance to illustrate our main results. In this section we restrict to perfect thermalization strokes at the end of which the working substance system reaches the Gibbs state. We relax this assumption in the subsequent Sec.~\eqref{sec:Finite thermalisation} and show that our central results are valid even with thermalisation strokes of finite duration. We summarize our findings and conclude in Sec.~\eqref{sec:Conclusion}. For the sake of completeness we summarize some of the known results that we have used in appendices \eqref{app:A} and \eqref{app:B}. In Appendix \eqref{app:C} we show that the results we have presented for average work output and reliability can also be extended to efficiency and its fluctuations. 

\section{\label{sec:Model}Asymmetric Quantum Otto Engine: Set-up}
The model for the asymmetric quantum Otto engine we consider here is detailed in \cite{PhysRevE.94.012137}. Briefly, we begin with the system in the state $\hrho_1$ at time $t_1$. The first compression work stroke from $t_1\leq t \leq t_2$ is a unitary realised by a time-dependent hamiltonian for the system of the form $\Hop[\omega_I(t)]$ with $\omega_I(t)$ denoting the time dependent control parameter defining the work protocol. Let us denote $\omega_I(t_1) = \omega_1$ and $\omega_I(t_2) = \omega_2$. Following this the system comes into contact with a hot bath with inverse temperature $\beta_h$ during the first heat exchange stroke from $t_2 \leq t \leq t_3$ (we set $\hbar = k_B = 1$ throughout). The cycle is completed by a second expansion work stroke with a time-dependent hamiltonian $\Hop[\omega_{II}(t)]$, with $\omega_{II}(t)$ giving the work protocol from $t_3 \leq t \leq t_4$, followed by a second heat exchange stroke realised by coupling the system to a cold bath at inverse temperature $\beta_c$ from between $t_4 \leq t \leq t_5$. During the heat exchange strokes the hamiltonian of the system is fixed. For cyclic operation we require that $\omega_{II}(t_4) = \omega_{I}(t_1) = \omega_1$. We can characterise the two work strokes via the unitary time evolution operators $\Uop_{\mathrm{WS}I}$ and $\Uop_{\mathrm{WS}II}$ defined via:
\begin{align}
    \Uop_{\mathrm{WS}I} &= \mathcal{T} e^{-i \int_{t_1}^{t_2} ds \Hop[\omega_I(s)]},\\
    \Uop_{\mathrm{WS}II} &= \mathcal{T} e^{-i \int_{t_3}^{t_4} ds \Hop[\omega_{II}(s)]},
\end{align}
with $\mathcal{T}$ denoting the chronological time-ordering operator. We model the dissipative heat exchange strokes in the limit of weak system-bath coupling by a Gorini-Kossakowski-Sudarshan-Lindblad (GKSL) Markovian master equation of the form \cite{Gorini1976,Lindblad1976}:
\begin{align}
\label{eq:master}
\frac{d\hat{\rho}}{dt} = -i\left[\Hop[\omega_i],\hat{\rho}\right ] + \mathcal{L}_{\beta_{\alpha}}\left[\hat{\rho}\right],   
\end{align}
with $i=2,\alpha = h$ ($i=1, \alpha=c$) denoting the heat stroke with the system coupled to the hot bath (cold bath). We will specify the jump operators and the exact form of the dissipative Liouvillian $\mathcal{L}_{\beta_{\alpha}}$ while discussing specific example systems for the working substance in Sec.~\ref{sec:Finite thermalisation}. Here, we note that in general the time-evolution during the heat strokes generated via the master equation Eq.~\eqref{eq:master} can be described via the completely positive maps $\hat{\Phi}_{\beta_h,t_3-t_2}[\cdot]$ and $\hat{\Phi}_{\beta_c,t_5-t_4}[\cdot]$. 

In typical realisations of the quantum Otto cycles considered in literature, such as \cite{Feldmann2004,Quan2007,PhysRevE.77.021128,Kosloff2017}, the first and the second work strokes are typically of the same duration $\tau_u/2$ and time-reversed versions of each other \emph{i.e.} $\omega_{I}(t) = \omega_{II}(\tau_u/2 - t)$. In the same manner, the dissipative strokes are also taken to have the same duration $\tau_b/2$. In contrast for the a-QOE cycle \cite{PhysRevE.94.012137}, the total time for the unitary stroke $\tau_u$ is divided into $t_2-t_1 = r_u\tau_u$ for the compression work stroke and $t_4-t_4 = (1-r_u)\tau_u$ for the expansion work stroke. In a similar manner the total time for the dissipative heat exchange strokes $\tau_b$ is divided into $r_b\tau_b$ for the thermalization with the hot bath and $(1-r_b)\tau_b$ for the thermalization with the cold bath. Indeed, our main focus in this paper is to optimize the average work and reliability as a function of the asymmetry parameters $r_u$ and $r_b$.


In order to extract the work statistics for the a-QOE cycle, we introduce projective measurements of the energy at the beginning and completion of the work strokes. This yields a sequence of four energies, denoted by $\{\epsilon_n^{(1)},\epsilon_m^{(2)},\epsilon_k^{(2)},\epsilon_l^{(1)}\}$, that characterize the system's energy profile during the cycle. Here $\epsilon_n^{(1)}$ and $\epsilon_l^{(1)}$ are the results of projective measurements at the beginning of the compression and at the end of expansion strokes, whereas $\epsilon_m^{(2)}$ and $\epsilon_k^{(2)}$ are the results of measurements at the end of the compression and at the beginning of the expansion strokes. These energies, as the superscript indices indicate, are given by the eigenvalues of either $\Hop[\omega_1]$ or $\Hop[\omega_2]$. The corresponding energy eigen-projectors $\Piop(j;\omega_i) = \ket{j;\omega_i}\bra{j;\omega_i}$ are defined by:
\begin{align}
    \Hop[\omega_i] \Piop(j;\omega_i)  = \epsilon_j^{(i)} \Piop(j;\omega_i),
\end{align}
with index $i=1,2$. For a given sequence of measurement results, the value of work done on the system is given by $w = (\epsilon_m^{(2)} - \epsilon_n^{(1)} + \epsilon_l^{(1)} - \epsilon_k^{(2)})$. In fact such a sequence of measurements also gives the heat exchanged, $q_h = \epsilon_k^{(2)} - \epsilon_m^{(2)}$, with the hot bath in the thermalization stroke that occurs between the work strokes. For the purposes of this paper, since we mainly focus on the work statistics, we can simply consider the marginal of the joint distribution of function of work and heat \emph{i.e.} the work distribution function that is given by \cite{PhysRevE.98.042122}:
 \begin{align}
 \label{eq:joint probability distribution}
 p(w) = \sum_{nmkl} \delta [w - (\epsilon_m^{(2)} - \epsilon_n^{(1)} + \epsilon_l^{(1)} - \epsilon_k^{(2)})]  p_{(4)}(l,k,m,n).
\end{align}
Here, $p_{(4)}(l,k,m,n)$ is the four-point conditional probability given by:
\begin{align}
 \label{eq:four-point conditional probability}
  p_{(4)}(l,k,m,n) = T_{II}(l,k)T_{\beta_h}(k,m)T_{I}(m,n)p_1(n), 
\end{align}
with the transition probabilities for the compression, expansion and dissipative stroke with the hot bath given respectively by:
\begin{align}
T_{I}(m,n) &= |\langle m;\omega_2| \Uop_{\mathrm{WS}I} | n;\omega_1 \rangle|^2 \label{eq:transprobcompress}\\
T_{II}(l,k) &= |\langle l;\omega_1| \Uop_{\mathrm{WS}II} | k;\omega_2 \rangle|^2 \label{eq:transprobexpand}\\
T_{\beta_h}(k,m) &= \Tr\left ( \Piop(k;\omega_2)\hat{\Phi}_{\beta_h,r_b \tau_b}\left[\Piop(m;\omega_2) \right] \right ) \label{eq:transhotbath}.
\end{align}
Finally, $p_1(n) = \Tr[\Piop(n;\omega_1) \hrho_1]$ is the population in the energy eigenbasis of the initial state $\hrho_1$. Anticipating the discussion of the finite time thermalization in Sec.~\ref{sec:Finite thermalisation}, we note that similar to Eq.~\eqref{eq:transhotbath}, we can also define a transition matrix for the final heat stroke arising from coupling the system to the cold bath as:
\begin{align}
    T_{\beta_c}(n^{\prime},l) &= \Tr\left(\Piop(n^{\prime};\omega_1)\hat{\Phi}_{\beta_c,(1-r_b) \tau_b}\left[\Piop(l;\omega_1) \right] \right) \label{eq:transcoldbath}.
\end{align}
In the limit $\tau_b \rightarrow \infty$ the system thermalizes perfectly with the bath and we have that,
\begin{align}
    T_{\beta_h}(k,m) = p_{\beta_h}(k) = \frac{e^{-\beta_h \epsilon_k^{(2)}}}{Z_{\beta_h}},\label{eq:transperfectherm},
\end{align}
with $Z_{\beta_h} = \Tr e^{-\beta_h \Hop[\omega_2]}$ the partition function of the system in equilibrium with the hot bath. By the same token, in this perfect thermalization limit the initial state produced after the heat stroke with the cold thermal bath will also be a Gibbs state \emph{i.e} $\hrho_1 = e^{-\beta_c \Hop[\omega_1]}/Z_{\beta_c}$, which leads to the following expression for $p_1(n)$ in Eq.~\eqref{eq:four-point conditional probability}:
\begin{align}
    p_1(n) = \frac{e^{-\beta_c \epsilon_k^{(1)}}}{Z_{\beta_c}}
    \label{eq:p1perfectherm},
\end{align}
with $Z_{\beta_c} = \Tr e^{-\beta_c \Hop[\omega_1]}$. In contrast for finite time thermalization the transition probability will depend on both indices $k$ and $m$. We will present the exact form of the transition probabilities for specific systems in Sec.~\eqref{sec:System}. 
From Eq.~\eqref{eq:joint probability distribution} we can obtain the $p^{\mathrm{th}}$ moment of the work distribution in a simple manner as:
\begin{align}
 \langle w^p \rangle = \sum_{nmkl} & (\epsilon_m^{f} - \epsilon_n^i + \epsilon_l^{i} - \epsilon_k^f)^p T_{II}(l,k)T_{\beta_h}(k,m)\nonumber\\
 & T_{I}(m,n)p_1(n) \label{eq:moments}.
\end{align}
Alternatively, the moments can also be obtained by defining the characteristic function of work $w$ given by
\begin{align}
 \label{eq:moments genrating function}
  G_w(\alpha)= \int p(w) e^{i\alpha w} dw.
\end{align}
From the characteristic function, the moments can simply be extracted by evaluating derivatives with respect to $\alpha$ as:
\begin{align}
\label{eq:Ho moments by generating function}
   \langle w^p \rangle = \left . \frac{d^p G_w(\alpha)}{d(i \alpha )^p} \right \vert_{\alpha=0}
\end{align}
The two key quantities of interest that characterize the engine are the average work output $-\avg{w}$, and a normalized measure of the fluctuation in the work output known as reliability of the engine $R_w$ defined as:
\begin{align}
    R_w = \frac{-\avg{w}}{\sigma_w} \label{eq:reliability},
\end{align}
with $\sigma_w = \sqrt{\avg{w^2}-\avg{w}^2}$ denoting the standard deviation of work. The central aim of our work is to examine the behaviour of $-\avg{w}$ and $R_w$ for the a-QOE as a function of $r_u$ and $r_b$ which parametrize the asymmetry in the duration of the strokes of the cycle. More specifically, we are interested in the optimization of the engine cycle operation as a function of $r_u$ such that the work output and reliability are maximised. This goes beyond the consideration in \cite{PhysRevE.94.012137} where the behaviour of the optimal value of $r_u$ and $r_b$ such that the average work output and efficiency is maximised was studied in detail. As we show in the forthcoming sections with specific systems, we find several interesting results. Perhaps the most remarkable one among them is the fact that for a given total protocol time $\tau_u+\tau_b$, there are some specific choices of the parameters $r_u$ characterizing the asymmetry of the protocol such that both the work output and reliability are maximised. Such \emph{co-optimal} values of the asymmetry $r_u$ are advantageous operating points where the engine provides a large and reliable work output.
\section{\label{sec:System} Asymmetric Quantum Otto Engine: Specific Systems}
In this section we sequentially consider two specific realization of the a-QOE with a harmonic oscillator (HO) and two-level system (TLS) as the working substance. For the sake of simplicity, we will restrict the discussion to the limit of perfect thermalization ($\tau_b \rightarrow \infty$) during the heat exchange strokes. This limit also allows us to obtain analytical expressions for the work output and reliability using well-known results regarding the unitary dynamics of HO \cite{PhysRevE.77.021128} and TLS systems \cite{e22111255}.

\subsection{\label{subsec:HO}Harmonic Oscillator Working Substance}
\begin{figure}
	\centering
	\subfloat{\begin{overpic}[width=0.9 \linewidth]{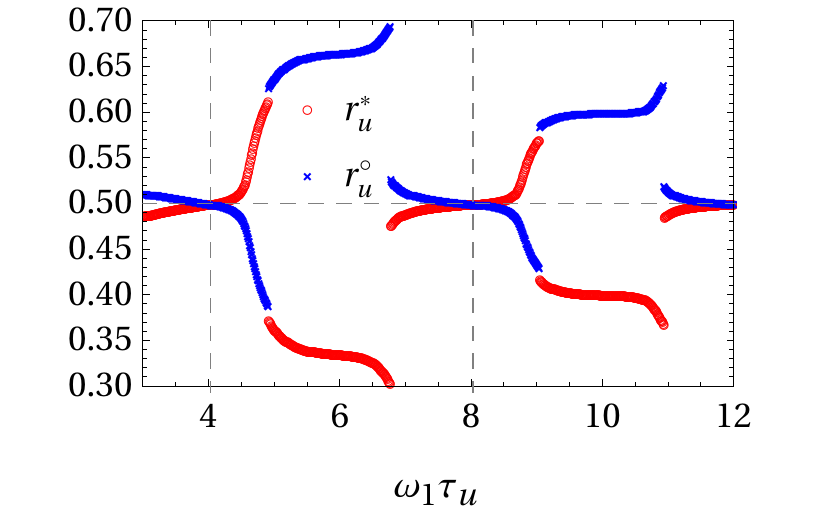}
	\put(20,50){\textbf{(a)}}
	\end{overpic}
	}
	\\ 
	\subfloat{\begin{overpic}[width=0.9 \linewidth]{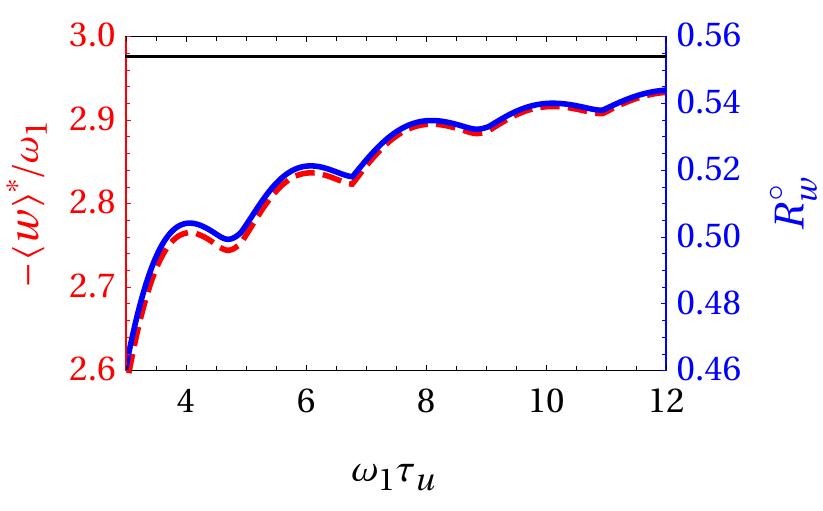}
	\put(20,50){\textbf{(b)}}
	\end{overpic}
	}\\ 
    \caption{(Color Online) Optimum values of the asymmetry parameters $r^*_{u}$ (red circles) and $r^{\circ}_{u}$ (blue crosses) (a) along with the corresponding values of average work output $-\langle w \rangle^*$ (red dashed line, left axis) and reliability $R^{\circ}_w$ (blue solid line, right axis) (b) as a function of total time of the cycle $\tau_u$ for an a-QOE with a harmonic oscillator as the working substance. The vertical dashed lines in (a) indicate the times for which the average work and reliability are co-optimized. The horizontal solid line in (b) indicates the optimal values in the quasi-static limit.  Parameter values are $\omega_2 = 2.0 \omega_1$, $\beta_h \omega_1 = 0.1$ and $\beta_c \omega_1 = 0.5$.}
	\label{fig:optimisationHO_wavgRwoptimise1}
\end{figure}
\begin{figure}
	\centering
	\subfloat{\begin{overpic}[width=0.9 \linewidth]{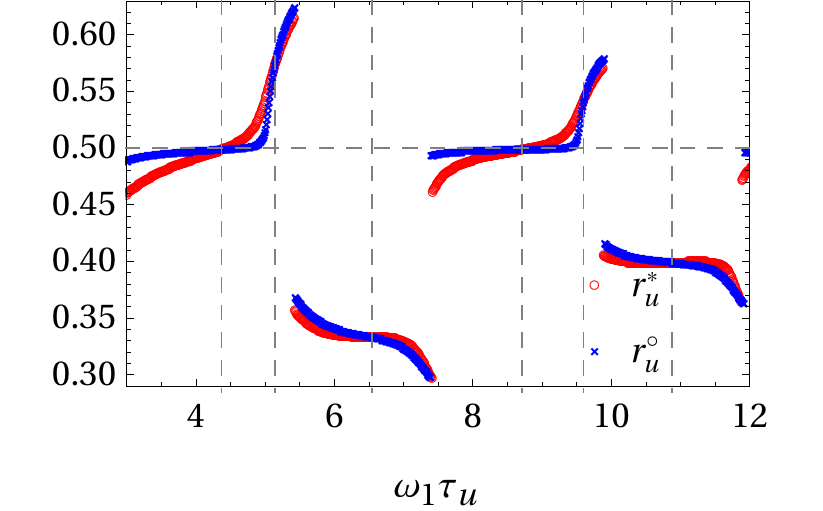}
	\put(20,50){\textbf{(a)}}
	\end{overpic}
	}\\ 
	\subfloat{\begin{overpic}[width=0.9 \linewidth]{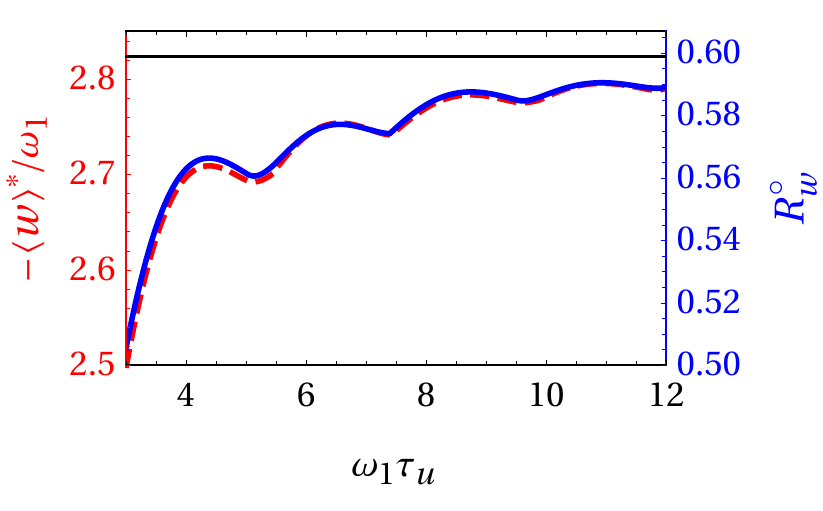}
	\put(20,50){\textbf{(b)}}
	\end{overpic}
	}\\ 
	\caption{(Color Online) Optimum values of the asymmetry parameters $r^*_{u}$ and $r^{\circ}_{u}$ (a) along with the corresponding values of average work output $-\langle w \rangle^*$ (red dashed line, left axis) and reliability $R^{\circ}_w$ (blue solid line, right axis) (b) as a function of total time of the cycle $\tau_u$ for cycle $\tau_u$ for an a-QOE with a harmonic oscillator as the working substance. The vertical dashed lines in (a) indicate the times for which the average work and reliability are co-optimized. The horizontal solid line in (b) indicates the optimal values in the quasi-static limit. Parameter values are  $\omega_2 = 1.8 \omega_1$, $\beta_h \omega_1 = 0.1$ and $\beta_c \omega_1 = 0.5$.}
	\label{fig:optimisationHO_wavgRwoptimise2}
\end{figure}

As the first example system we consider a harmonic oscillator whose time-dependent angular frequency $\omega(t)$ serves as the control parameter during the work strokes. The hamiltonian in this case is given by:
\begin{align}
\label{eq:HOham}
 \Hop_{\mathrm{HO}}[\omega(t)] &= \omega(t)\left( \hat{n}(t)+ \frac{1}{2} \right),
\end{align}
where the time-dependent number operator is given by $\hat{n}(t) = \adop(t) \aop(t)$ with $\aop(t) = \sqrt{m\omega(t)/2} [\,\xop + i \pop/(m\omega(t))\,]$. As described in Section.~\eqref{sec:Model}, during the the work strokes the angular frequency is swept from $\omega(t_1) = \omega_1$ to $\omega(t_2) = \omega_2$. With $\omega_2>\omega_1$, the forward and reverse work strokes can precisely be understood as compression and expansion strokes respectively. Interestingly for the HO, in the perfect thermalization limit ($\tau_b \to \infty$), the all important characteristic function Eq.~\eqref{eq:moments genrating function} and hence the moments can be written in terms of two parameters $Q_f$ and $Q_b$ that encode the details of the work protocol. The parameters $Q_f$ and $Q_b$ characterize how close to quasi-static (or `quantum' adiabatic) the compression and expansion work strokes respectively are \cite{PhysRevE.77.021128} and depend parametrically on $r_u$ and $\tau_u$. We present the expressions for $Q_f$ and $Q_b$ and the rather cumbersome analytical expressions for the gennerating function in Appendix.~\eqref{app:A}. 

Before we consider the exact form of the work protocol and present the results, we want to comment on some of the limiting case scenarios. In the limit of a perfect quasi-static compression and expansion work protocol \emph{i.e.} $\tau_u \rightarrow \infty$, the asymmetry parameter $r_u$ plays no role and we have that $Q_f = Q_b \rightarrow 1$. In this limit the work output and reliability are obtained as:
\begin{align}
    -\avg{w} =& \frac{(\omega_2-\omega_1)}{2}\left[ \coth\left(\frac{\beta_h\omega_2}{2}\right)-\coth\left(\frac{\beta_c\omega_1}{2}\right)\right] \label{eq:HOquasistat_work}\\ 
    R_w =& \frac{(\omega_2 - \omega_1)^2\left[\cosech^2\left(\frac{\beta_c\omega_1}{2}\right)+\cosech^2\left(\frac{\beta_h\omega_2}{2}\right)\right]}{-\avg{w}}
  \label{eq:HOquasistat_reliability}.
\end{align}

We note that this is the global optimum operation point for the quantum Otto heat engine with HO in the sense that any of the finite time protocols, asymmetric or symmetric, will have lesser work output and reliability \cite{Plastina2014,Alecce2015}. Nonetheless, since the absolute quasi-static limit is practically unattainable it is important to understand how best to optimize the finite time cycle. It is in this context that Zheng et.al. \cite{PhysRevE.94.012137} found that for quantum Otto engine with a finite duration of work strokes $\tau_u$, symmetric protocol with $r_u=0.5$ is not necessarily the optimum choice to obtain maximum work output for every $\tau_u$. With this as the jumping-off point, we augment the results in \cite{PhysRevE.94.012137} by examining the optimization of the reliability of the work output in addition to its average value.

For purposes of illustrating our results, we restrict to a simple linear protocol of the form $\omega_{I}^2(t) = \omega_1^2 + (\omega_2^2 -\omega_1^2)(t-t_1)/(r_u\tau_u)$ and $\omega_{II}^2(t) = \omega_2^2 + (\omega_1^2 -\omega_2^2)(t-t_3)/((1-r_u)\tau_u)$ for the compression and expansion work strokes respectively. With this, at each value of the total protocol duration $\tau_u$, we find the optimal values of the asymmetry parameter $r_u = r^*_u$ and $r_u = r^{\circ}_u$ that maximise the work output and reliability respectively. In Figs.~\eqref{fig:optimisationHO_wavgRwoptimise1} and \eqref{fig:optimisationHO_wavgRwoptimise2} (a) we see that, in agreement with the central findings in \cite{PhysRevE.94.012137}, $r^*_{u}$ (depicted by red circles) is a discontinuous function of the duration of the protocol $\tau_u$ and in general, except for some isolated values of $\tau_u$, the symmetric protocol with $r_u = 0.5$ is not the one maximising work output. In addition Figs.~\eqref{fig:optimisationHO_wavgRwoptimise1} and \eqref{fig:optimisationHO_wavgRwoptimise2} (b) show that the absolute maximum value of the optimum work output (dashed red line) is always lesser than the quasi-static work (solid horizontal line). For the sake of completeness we also note that the smallest work output is obtained in the limit of $\tau_u \rightarrow 0$ \emph{i.e.} an instantaneous quench protocol. In this case the work output can be computed from observing that the parameter $Q_f = Q_b \to \frac{\omega_2^1+\omega_1^2}{2\omega_2\omega_1}$ in the limit of $\tau_u \to 0$. Additionally, in Figs.~\eqref{fig:optimisationHO_wavgRwoptimise1} and \eqref{fig:optimisationHO_wavgRwoptimise2} (a), we see that the optimum value $r_u^\circ$ (depicted by blue crosses) maximising the reliability also has discontinuities like $r_u^*$. Note however that the optimum values of $r_u$ for the work output and reliability are in general not the same in general. Moreover the optimum reliability $R_w^\circ$ (solid blue line) asymptotically tends to the adiabatic value [solid horizontal line in Figs.~\eqref{fig:optimisationHO_wavgRwoptimise1} \eqref{fig:optimisationHO_wavgRwoptimise2} (b)] as $\omega_1 \tau_u \gg 1$. Thus, we see that the discontinuities pointed out in \cite{PhysRevE.94.012137} for the work output also persist for the reliability. As described there, this discontinuity can be understood by the fact that $-\avg{w}$ and $R_w$ are non-monotonic functions of the asymmetry $r_u$ (at fixed $\tau_u$) with multiple local extrema. As $\tau_u$ increases, the identity of which local minima becomes the global minima can change suddenly leading to the discontinuities observed here. 

Since an advantageous regime of operation of the engine would be one where both the work output and reliability are as large as they can be, we ask if there are values of the engine cycle time $\tau_u$ such that the optimal $r_u$ for work output and reliability coincide. This is answered in the affirmative as shown by the vertical dashed lines in Figs.~\eqref{fig:optimisationHO_wavgRwoptimise1} and \eqref{fig:optimisationHO_wavgRwoptimise2} (a). This is one of our central results. For the parameters chosen in Fig.~\eqref{fig:optimisationHO_wavgRwoptimise1} (a) the co-optimal points of operation (vertical dashed lines indicating $\tau_u$) are obtained for symmetric protocols with $r_u^* = r_u^{\circ} = 0.5$. In contrast for a different set of parameters shown in Fig.~\eqref{fig:optimisationHO_wavgRwoptimise2} (a) we obtain co-optimisation at asymmetric protocol with $\left \{ r_u^* = r_u^{\circ} \right \} \neq 0.5$.

\subsection{\label{subsec:TLS}Working substance consists of TLS}
\begin{figure}
	\centering
	\subfloat{\begin{overpic}[width=0.9 \linewidth]{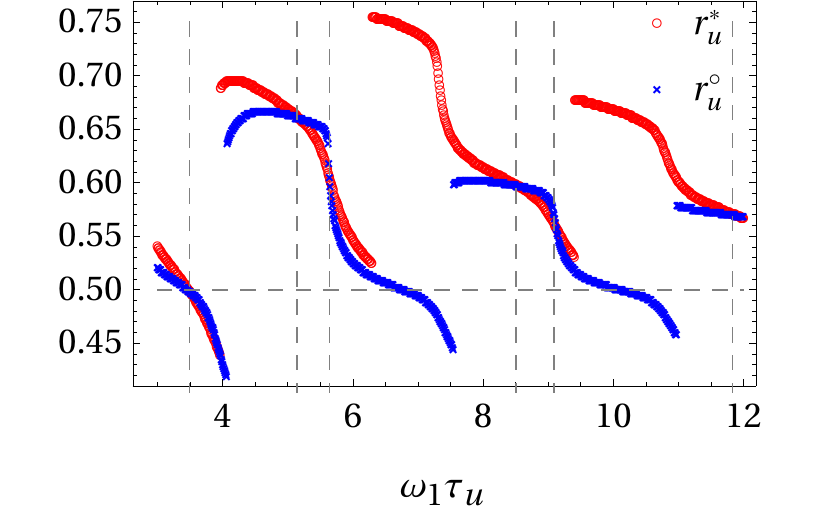}
	\put(18,50){\textbf{(a)}}
	\end{overpic}
	}\\ 
	\subfloat{\begin{overpic}[width=0.9 \linewidth]{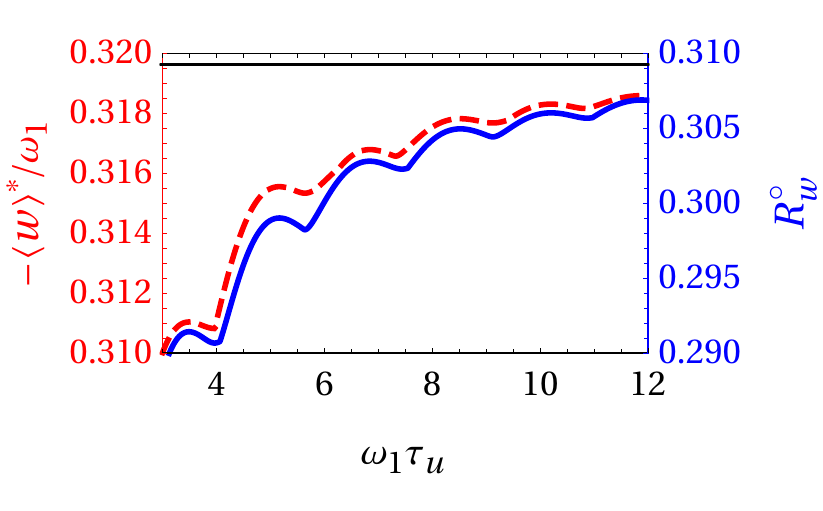}
	\put(20,50){\textbf{(b)}}
	\end{overpic}
	} 
	\caption{(Color Online) Optimum values of the asymmetry parameters $r^*_{u}$ and $r^{\circ}_{u}$ (a) along with the corresponding values of average work output $-\langle w \rangle^*$ (red dashed line, left axis) and reliability $R^{\circ}_w$ (blue solid line, right axis) as a function of total time of the cycle $\tau_u$ for an a-QOE with a two-level system as the working substance. The vertical dashed lines in (a) indicate the times for which the average work and reliability are co-optimized. The horizontal solid line in (b) indicates the optimal values in the quasi-static limit. Parameter values are $\omega_2 = 2.0 \omega_1$, $\delta = \omega_1$, $\beta_h \omega_1 = 0.1$ and $\beta_c \omega_1 = 0.5$.}
	\label{fig:optimisationTLS_wavg}
\end{figure}

Having demonstrated our central results with the HO system, we now are interested in understanding their generality. Since a system independent analytic approach for the a-QOE is not tractable, we explore this question by considering a second working substance system consisting of a two-level system (TLS) with the following time-dependent hamiltonian \begin{align}
\label{eq:TLS Hamiltonian}
  \Hop_{\mathrm{TLS}}[\omega(t)] &= \omega(t)\sz + \delta \sx,
\end{align}
with $\sz$ and $\sx$ the Pauli $x,z$ matrices. Thus the work strokes are actuated by sweeping the coefficient of the $\sz$ term $\omega(t)$ making this an effective Landau-Zener type problem. As in the HO case, the characteristic function for arbitrary protocols $\omega(t)$ can be written in terms of a single parameter $Q$ that takes the values $Q_f$ and $Q_b$ for the compression and expression strokes. The parameter $Q$ in this case simply denotes the probability of transition from the initial energy eigenstate in the TLS case and thus has a different interpretation from the parameter introduced for the HO (see Appendix.~\ref{app:A2}). For illustrating our results, in order to keep the description comparable to the HO, we sweep $\omega^2(t)$ in a linear manner with the protocols having the exact same form as in the previous sub-section. 

Fig.~\eqref{fig:optimisationTLS_wavg} shows that the discontinuity in the optimal value of $r_u$ for the work output discussed in \cite{PhysRevE.94.012137} for a HO system also extends to the TLS case. Moreover the extension of this result to the reliability discussed for the HO in the previous sub-section also carries through. Finally, our result that for particular engine cycle times $\tau_u$ we have co-optimization of work output and reliability with $r_u^* = r_u^\circ$ is also valid for the TLS system as illustrated clearly by the vertical dashed lines in Fig.~ \eqref{fig:optimisationTLS_wavg} (a). Comparing Figs.~\eqref{fig:optimisationHO_wavgRwoptimise1}, \eqref{fig:optimisationHO_wavgRwoptimise2}, and \eqref{fig:optimisationTLS_wavg} we see that for the TLS working substance the values of $\tau_u$ at which the optimal $r_u$ for the average work and reliability have discontinuities do not match up unlike the case of the HO. In fact we have verified that in the TLS case the discontinuities in $r_u^\circ$ coincide with the points of discontinuity in the values of $r_u$ that minimizes the fluctuation $\sigma_w$.

Having illustrated our central results for two disparate working substances in HO and TLS, we next proceed to generalize further and consider a-QOE cycles with thermalization strokes with finite duration $\tau_b$. 
\section{\label{sec:Finite thermalisation} Finite Time Thermalization}
The probability to obtain the result $\epsilon^{(1)}_{n'}$ for an energy measurement on the working system at the end of the four strokes given that the result of the initial measurement was $\epsilon^{(1)}_{n}$ is given by the cycle transition matrix \cite{PhysRevE.94.012137,PhysRevE.98.042122}:
\begin{align}
  T_{\mathrm{cyc}}(n^\prime,n) & = \sum_{l,k,m} T_{\beta_c}(n^\prime,l)T_{II}(l,k) \nonumber \\
  & T_{\beta_h}(k,m)T_{I}(m,n)p_1(n) \label{eq:Tcycle Otto}.
\end{align}
When both the heat and work strokes are of finite duration, the cycle transition matrix depends both on the initial state $\hrho_1$ of the cycle and the details of the time evolution during the thermal stroke given by Eq.~\eqref{eq:master}. This is in contrast with the infinite thermalization limit where the cycle transition matrix becomes trivial and independent of $n$ i.e. $T_{\mathrm{cyc}}(n^\prime,n) = p_{\beta_c}(n')$. Clearly, in this case the cycle is naturally closed in the sense that the working substance returns to the same state automatically after each cycle. Thus, our first consideration with the finite time thermalization scenario is to identify an initial state for the working substance. A natural choice emerges from the observation that the cycle transition matrix $T_{\mathrm{cyc}}$ is irreducible and hence we can find an eigenvector with eigenvalue $1$ \emph{i.e.} a stationary state for the cycle. In the following discussion of the a-QOE with finite time thermalization, we take the initial state probability distribution as this properly normalized eigenvector with eigenvalue $1$ of the cycle transition matrix.
\begin{figure}
	\centering
	\subfloat{\begin{overpic}[width=0.9 \linewidth]{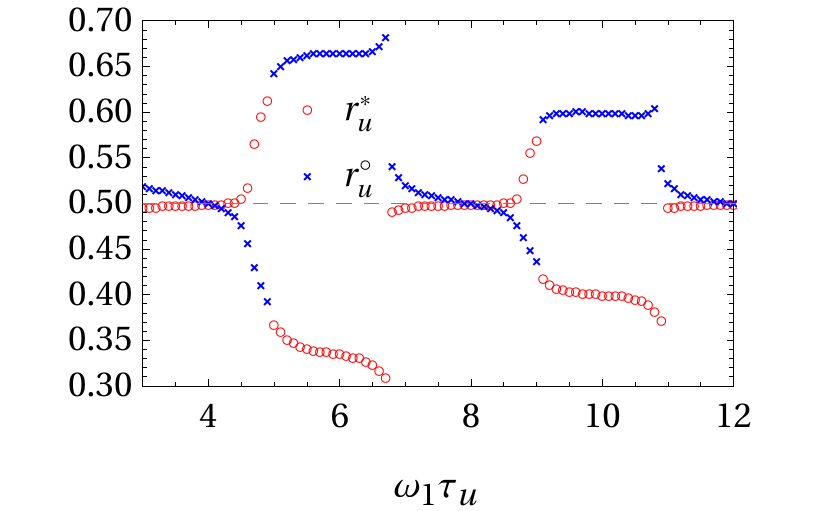}
	\put(20,50){\textbf{(a)}}
	\end{overpic}
	}\\ 
	\subfloat{\begin{overpic}[width=0.9 \linewidth]{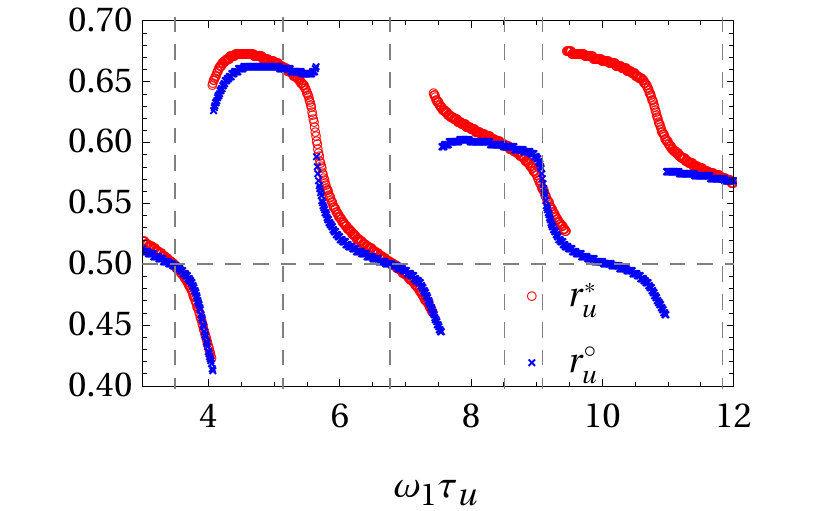}
	\put(19,50){\textbf{(b)}}
	\end{overpic}
	} 
	\caption{(Color Online)  \label{fig:co-optimiation finite thermalisation} Co-optimum values of the asymmetry parameters  ($r^*_{u}$ and $r^{\circ}_{u}$) as a function of total cycle time $\tau_u$ for HO (a) and TLS (b) working substance with finite time thermalisation strokes. Parameter values are $\omega_2 = 2.0 \omega_1$, $\beta_h \omega_1 = 0.1$ and $\beta_c \omega_1 = 0.5$ for the HO and $\omega_2 = 2.0 \omega_1$, $\delta = \omega_1$, $\beta_h \omega_1 = 0.1$ and $\beta_c \omega_1 = 0.5$ for the TLS. For the thermalization strokes, $r_b=0.5$ and $\tau_b = 10 \tau_u$ is chosen in both cases.}
\end{figure}
In a GKLS master equation, the general form of the Liouvillian representing dissipation via an arbitrary jump operator $\hO$ is given by:
\begin{align}
    \mD_{\hO}[\hrho] = \left(2 \hO \hrho \hO^{\dagger} - \hO^{\dagger}\hO \hrho - \hO^{\dagger}\hO \hrho  \right) \label{eq:jumpliouv}.
\end{align}
With this we can write down the dissipative part of the Liouvillian for the master Eq.~\eqref{eq:master} representing the interaction of the HO with a thermal bath:
\begin{align}
    \mL_{\beta_i}^{\mathrm{HO}}[\hrho] = \kappa (\bar{n}(\omega_i)+1)\mD_{\aop_i} [\hrho] +  \kappa \bar{n}(\omega_i)\mD_{\adop_i} [\hrho]. \label{eq:Ho dissipative master equation}
\end{align}
Here $\kappa$ is the damping rate, $\bar{n}(\omega_i) = e^{-\beta_i \omega_i}/(1-e^{-\beta_i\omega_i})$ is the thermal occupation number of the bath,  $\aop_i$ denotes the ladder operator for the HO system with frequency $\omega_i$, and $i=1,2$ with $\beta_1 = \beta_c,\beta_2 = \beta_h$. Finally, in order to stay within the weak-coupling regime that ensures the validity of the GKLS master equation, we have to always ensure that $\kappa (\bar{n}(\omega_i)+1)/\omega_i \ll 1$ \cite{PhysRevE.94.012137}. In an analogous manner we can write down the dissipative part for the TLS as:
\begin{align}
    \mL_{\beta_i}^{\mathrm{TLS}}[\hrho] = \gamma (\bar{m}(\omega_i)+1)\mD_{\hs_i^-} [\hrho] +  \gamma \bar{m}(\omega_i)\mD_{\hs_i^+} [\hrho]. \label{eq:TLS dissipative master equation}
\end{align}
Here the operator $\hs_i^- = \ket{-,\omega_i} \bra{+,\omega_i}$ is the spin ladder operator with $\ket{\pm,\omega_i}$ denoting the eigenstates of the TLS hamiltonian \emph{i.e.} $\Hop_{\mathrm{TLS}}[\omega_i] \ket{\pm,\omega_i} = \pm \sqrt{\omega_i^2+\delta^2} \ket{\pm,\omega_i}$. The bath occupation number is given by $\bar{m}(\omega_i) = 1/(e^{2 \beta_i \sqrt{\omega_i^2+\delta^2}}+1)$. As in the HO case, $i=1,2$ and $\beta_1=\beta_c,\beta_2=\beta_h$ and we have to ensure that  $\gamma (\bar{m}(\omega_i)+1)/\sqrt{\omega_i^2+\delta^2} \ll 1$ for the validity of weak coupling.

Having written down the form of the dissipative Liouvillian, we can proceed and calculate the work output and reliability for the finite thermalization a-QOE with the HO and TLS systems. Considering the HO system first, we detail the steps involved in such a calculation as opposed to the infinite time thermalization presented in Sec.~\eqref{subsec:TLS}. The key step is to calculate the eigenvector with eigenvalue $1$ of the cycle transition matrix Eq.~\eqref{eq:Tcycle Otto}. In order to do this we need the transition matrix elements for both the unitary and dissipative strokes that appear in Eq.~\eqref{eq:Tcycle Otto} in the HO energy basis. For the sake of completeness we provide these well-known expressions \cite{PhysRevE.98.042122} in Appendix B. Note that unlike the infinite thermalization case where we were able to calculate the characteristic function analytically without appeal to the HO energy basis, in this case we have to numerically evaluate the cycle transition matrix with a cut-off on the number of eigenstates of the oscillator. In contrast, for the TLS case, even for the finite time thermalization we can write down analytical expressions (though cumbersome) for the cycle transition matrix and also diagonalize the same to identify the initial state for the cycle (see Appendix B). In both the cases, the dissipative dynamics of the master equation reduces to a simple rate equation for the purposes of calculating the transition matrices $T_{\beta_i}(k,m)$. Once the initial state probability distribution $p_1(n)$ is calculated, we can proceed to obtain the work output and reliability using Eq.~\eqref{eq:moments}.

In Fig.~\eqref{fig:co-optimiation finite thermalisation} (a) we have illustrated the behaviour of the optimal values $r_u^*$ and $r_u^\circ$ for the HO case with finite time thermalization strokes for the same parameters as in Fig.~\eqref{fig:optimisationHO_wavgRwoptimise1}. Clearly we again have particular $\tau_u$ at which there is co-optimization. We have used a cut-off of $50$ oscillator eigenstates to obtain the results. Moreover we have chosen a somewhat coarser grid for $\tau_u$ in Fig.~\eqref{fig:co-optimiation finite thermalisation} (a) as the calculation of the cycle transition matrix is numerically intensive. In line with HO case, Fig.~\eqref{fig:co-optimiation finite thermalisation} (b) illustrates the co-optimization result for the TLS with finite time thermalization strokes. In both Figs.~\eqref{fig:co-optimiation finite thermalisation} (a) and (b), the bath stroke is taken to be symmetric \emph{i.e.} $r_b = 0.5$. While we have made this choice for simplicity, we have also checked that our results are qualitatively similar for other values of $r_b$. Thus, all of our central results from Sec.~\eqref{sec:System} such as discontinuity in the asymmetry of the work stroke duration ($r_u$) that optimize work output and reliability and co-optimization of work output and reliability are valid even with finite time thermalization.

\section{\label{sec:Conclusion}Conclusion}
In this article we have considered an quantum Otto cycle with asymmetric compression and expansion strokes. The asymmetry in the strokes was manifested by choosing different duration for the compression work strokes. We then optimized two figures of merit characterizing the engine performance, the average work output $-\avg{w}$ and the reliability $R_w$, as a function of the parameter $r_u$ characterizing the asymmetry of the work strokes for two specific working substance systems give by a harmonic oscillator and two-level systems. We showed that the result in \cite{PhysRevE.94.012137} demonstrating discontinuities in the optimal $r_u$ maximising work output extends to the reliability as well as the efficiency and its fluctuations (see App.~\eqref{app:C}). Moreover we showed that there are some values of the total time for the work strokes $\tau_u$ at which both the work output and its fluctuations characterized by the reliability are maximized for the same value of the asymmetry parameter \emph{i.e.} they are co-optimized. Finally, we also demonstrated that our central results of discontinuities and co-optimization are also valid when the thermalization strokes are taken to be of finite duration and modelled via a GKLS master equation. Apart from extending the validity of the results the finite thermalization study also allows to use the power output of the engine as a figure of merit equivalent to the work output. Hence the results for the finite thermalization indicate that when a finite time is available for the total engine cycle, it is often advantageous in terms of a higher average output power to distribute the time unequally between the compression and expansion strokes. For the sake of simplicity in the examples considered here, we have taken the asymmetry parameter for the thermal strokes $r_b = 0.5$. An interesting follow-up of this work is a systematic study of how the asymmetry in the bath strokes affects the power output and its fluctuations.

\begin{acknowledgments}
This work was supported and enabled by a MoE (India) funded Ph.D. fellowship at IIT Gandhinagar (R.~S.~) and the  Department of Science \& Technology Science and Engineering Research Board (India) Start-up Research Grant No. SRG/2019/001585 (B.~P.~V). We thank Bijay Agarwalla and Jayanth Jayakumar for many fruitful discussions.
\end{acknowledgments}
\appendix
\section{\label{app:A}Characteristic Function Calculation for Perfect Thermalization}
In this appendix, we summarize some known results that enable the calculation of work output and reliability for the infinite time/perfect thermalization limit presented in the main text. 

Anticipating the calculation of the efficiency in Appendix.~\eqref{app:C}, we consider the joint probability distribution function (pdf) of total work done $w$ and heat supplied $q_h$ for the Otto cycle introduced in the main text:
\begin{align}
   \label{eq:joint probability distribution}
p(w,q_h) =& \sum_{nmkl} \delta [w - (\epsilon_m^{(2)} - \epsilon_n^{(1)} + \epsilon_l^{(1)} - \epsilon_k^{(2)})] \nonumber\\
 &\times \delta [q_h - (\epsilon_k^{(2)} - \epsilon_m^{(2)})] p^{(4)}(l,k,m,n) 
\end{align}
The generating function corresponding to this joint pdf is defined as:
\begin{align}
\label{eq:moments genrating function Full}
G_{w,q_h}(\alpha,\bar{\alpha})= \int p(w,q_h) e^{i\alpha w} e^{i\bar{\alpha} q_h} dw dq_h.
\end{align}
Note that the characteristic function of the work distribution given by Eq.~\eqref{eq:moments genrating function} can be easily obtained as $G_w(\alpha) = G_{w,q_h}(\alpha,0)$. All the moments of the form $\langle w^n q_h^m \rangle$ can be written down by differentiating Eq.~\eqref{eq:moments genrating function Full} as:
\begin{align}
\label{eq:Ho moments by generating function}
   \langle w^p q_h^s \rangle = \left . \frac{\partial^p \partial^s G(\alpha,\bar{\alpha})}{\partial(i \alpha )^p\partial(i \bar{\alpha} )^s} \right \vert_{\alpha=0,\bar{\alpha}=0}.
\end{align}
With this, we present below the expressions for the moment generating function for the infinite thermalization case for the two working substance systems considered in the paper.
\subsection{HO Working Substance}\label{app:A1}
The generating function for the HO takes the form \cite{PhysRevE.77.021128},
\begin{align}
\label{eq:6}
 G^{\mathrm{HO}}_{w,q_h}(\alpha,\bar{\alpha}) = \frac{2e^{\frac{(\beta_c \omega_1+\beta_h\omega_2)}{2}}}{Z_{\beta_c}Z_{\beta_h}}f(Q_f,x_0,y_0)f(Q_b,x_1,y_1),
\end{align}
with the function
\begin{align}
\label{eq:7}
f(Q,x,y) =& \frac{1}{\sqrt{Q(1-x^2)(1-y^2) + (1+x^2)(1+y^2) -4xy }}\nonumber \\
\end{align}
and $x_0 = e^{-(i\alpha+\beta_c)\omega_1}$, $y_0 = e^{(i\alpha-i \bar{\alpha})\omega_2}$, $x_1 = e^{-(i\alpha-i\bar{\alpha}+\beta_h)\omega_2}$, and $y_1 = e^{i\alpha \omega_1}$. Here, as described in the main text, all information about the work protocol $\omega(t)$ is encoded in the function $Q(\tau_p,\omega_1,\omega_2)$ which is defined as follows. If $X(t)$ and $Y(t)$ are solutions of the classical equations of motion (EOM): $\Ddot{X} + \omega^2(t)X = 0$ with initial conditions $X(0) = 0$, $\dot{X}(0)=1$ and $Y(0) = 1$, $\dot{Y}(0)=0$ then,
\begin{align}
\label{eq:Q}
Q(\tau_p,\omega_1,\omega_2) & = \frac{1}{2\omega_1\omega_2}\left[\omega_1^2\left(\omega_2^2X(\tau_p)^2 + \dot{X}(\tau_p)^2\right)+ \right . \\
& \left .
\left(\omega_2^2Y(\tau_p)^2 + \dot{Y}(\tau_p)^2\right)\right], \nonumber
\end{align}
with $\tau_p$ denoting the protocol time over which the oscillator's frequency is modulated from $\omega_1$ to $\omega_2$. With this we identify $Q_f = Q(r_u \tau_u, \omega_1,\omega_2)$ and $Q_b = Q((1-r_u)\tau_u,\omega_2,\omega_1)$. In the results presented in the main text, we calculate the function $Q(\tau_p,\omega_1,\omega_2)$ by a numerical integration of the classical EOM.

\subsection{TLS Working Substance}\label{app:A2}
For the TLS working substance the characteristic function takes the form:
\begin{align}
\label{eq:perfect thermal generating function}
 G^{\mathrm{TLS}}_{w,q_h}(\alpha,\bar{\alpha}) &= \frac{1}{Z_{\beta_c}Z_{\beta_h}} 
 \left(Q_fx_- -(Q_f-1)x_+\right) \nonumber\\&\left(Q_by_--(Q_b-1)y_+ \right), 
\end{align}
with $x_{\pm} = \cos \left[(\alpha-\bar{\alpha})\Delta_2 + \pm (\alpha-i \beta_c) \Delta_1 \right]$, $y_{\pm} = \cos \left[\alpha \Delta_1 \pm (\alpha-\bar{\alpha}-i \beta_h) \Delta_2\right]$, and $\Delta_i = \sqrt{\delta^2+\omega_i^2}$. As in the HO case, we introduce the function $Q(\tau_p,\omega_1,\omega_2)$ that encodes the nature of the work protocol. Here, this function is related in a simple manner to the staying probability in the instantaneous energy eigenstates and is given by:
\begin{align}
    Q(\tau_p,\omega_1,\omega_2) = \vert \sandwich{\omega_2,\pm}{\mathcal{T} e^{-\frac{i}{\hbar} \int_{0}^{\tau_p} ds \Hop_{\mathrm{TLS}}[\omega(s)]}}{\omega_1,\pm} \vert^2.
\end{align}
With this, we have that the parameters in Eq.~\eqref{eq:perfect thermal generating function} can be written as as $Q_f = Q(r_u \tau_u,\omega_1,\omega_2)$ and $Q_f = Q((1-r_u) \tau_u,\omega_2,\omega_1)$. As in the HO case, we calculate the function $Q(\tau_p,\omega_1,\omega_2)$ by a numerical integration of the Schr\"{o}dinger equation for the TLS dynamics.\\
\vspace{-0.1in}

\section{\label{app:B}Transition matrix elements calculation for Finite Time Thermalization}
In this appendix, we summarize some known results that enable the calculation of the matrix elements for the unitary and dissipative strokes of a-QOE. This is required to evaluate the cycle transition matrix presented in Eq.~\eqref{eq:Tcycle Otto} in the main paper. 
\subsection{HO Working Substance}
The key to calculate the cycle transition matrix $T_{\mathrm{cyc}}$ in Eq.~\eqref{eq:Tcycle Otto} is to obtain the transition probability matrices on the right hand side of the equation such as $T_{\beta_c/\beta_h}$, $T_{I/II}$. To this end we note that for the work strokes we can write the transition matrix elements as \cite{PhysRevE.98.042122}:
\begin{widetext}
\begin{align}
\label{eq:Ho unitary transition probabilities}
   T_{J}(m,n) = \begin{cases} 
      \frac{\sqrt{2}}{\sqrt{Q_{J}+1}} \left( \frac{Q_{J}-1}{Q_{J}+1} \right)^{(m+n)/2} \frac{\Gamma\left[(m+1)/2\right]\Gamma\left[(n+1)/2\right]}{\pi \Gamma\left[m/2+1\right]\Gamma\left[n/2+1\right]} {}_2F_1\left( -\frac{m}{2},-\frac{n}{2};\frac{1}{2};\frac{2}{1-Q_{J}}\right)^2 & m,n : \text{even} \\
      \frac{2^{7/2}}{(Q_{J}+1)^{3/2}} \left( \frac{Q_{J}-1}{Q_{J}+1} \right)^{(m+n)/2-1} \frac{\Gamma\left[m/2+1\right]\Gamma\left[n/2+1\right]}{\pi \Gamma\left[(m+1)/2\right]\Gamma\left[(n+1)/2\right]} {}_2F_1\left( \frac{1-m}{2},\frac{1-n}{2};\frac{3}{2};\frac{2}{1-Q_{J}}\right)^2  & m,n : \text{odd} \\
      0 & \text{else}.
   \end{cases}
\end{align}
\end{widetext}
Here ${}_2F_1(a,b;c;z)$ denoting the hypergeometric function and the index $J$ takes the values $I$ ($II$) on the left hand side and $f$ ($b$) on the right hand side for the compression (expansion) strokes. The values $Q_f$ and $Q_b$ are protocol dependent and are as described in Appendix \ref{app:A1}. For the thermal strokes described by Eq.~\eqref{eq:Ho dissipative master equation} for a HO of frequency $\omega$, of duration $\tau$ in contact with a bath with temperature $\beta$, the transition probability in the energy basis can be written in general as \cite{PhysRevE.98.042122}:
\begin{align}
\label{eq:Ho finite time thermal transition probabilities}
 \bar{T}_{\beta}(n,l,\tau,\omega) =& \frac{1-\theta_{\tau}}{1-\nu \theta_{\tau}} \nu^n \sum_{i=0}^{\text{min}(n,l)} \frac{(-1)^i(n+l-i)!}{(n-i)!(l-i)!i!} \nonumber\\
 &\times\left( \frac{1-\theta_{\tau}}{1-\theta_{\tau} \nu} \right)^{n+l-i}\left( \frac{1-\frac{\theta_{\tau}}{\nu}}{1-\theta_{\tau}} \right)^i.
\end{align}
Here, $\theta_{\tau} = e^{-2 \kappa \tau}$ is a dimensionless parameter determining how far the thermalisation has proceeded within the duration $\tau$, and $\nu = e^{-\beta \omega}$. With this, we can write the relevant transition matrices in Eq.~\eqref{eq:Tcycle Otto} as $T_{\beta_c}(n^\prime,l) = \bar{T}_{\beta_c}(n^\prime,l,(1-r_b)\tau_b,\omega_1)$ and $T_{\beta_h}(k,m) = \bar{T}_{\beta_h}(k,m,r_b\tau_b,\omega_2)$.\\

\subsection{TLS Working Substance}
\begin{figure}
	\centering
	\subfloat{\begin{overpic}[width=0.9 \linewidth]{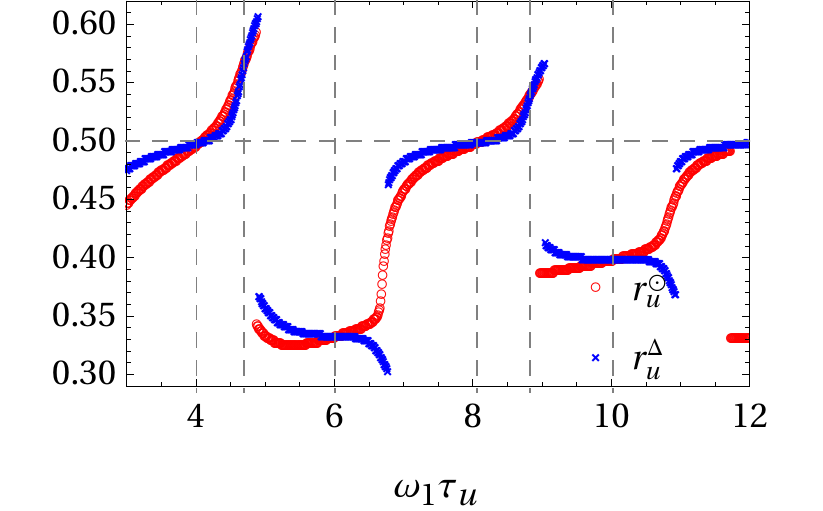}
	\put(20,50){\textbf{(a)}}
	\end{overpic}
	}\\
	\subfloat{\begin{overpic}[width=0.9 \linewidth]{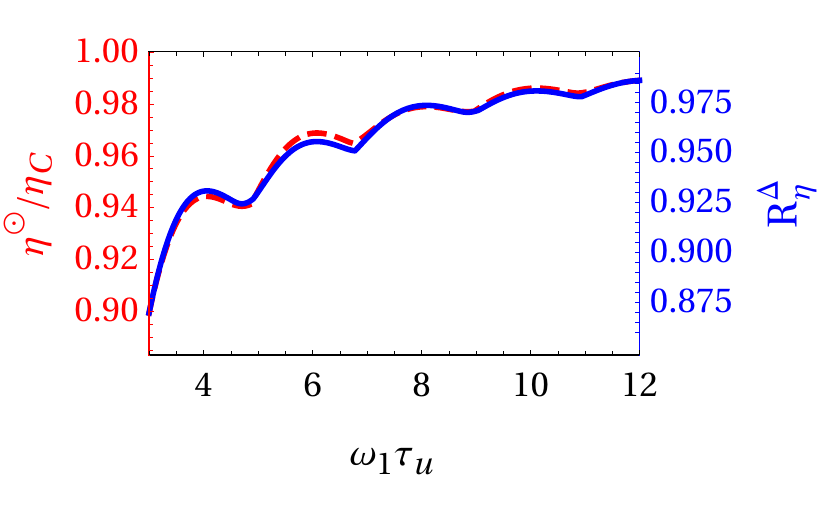}
	\put(20,50){\textbf{(b)}}
	\end{overpic}
	}
	\caption{(Color Online) Optimum values of the asymmetry parameters $r^{\odot}_{u}$ and $r^{\Delta}_{u}$ (a) along with the corresponding maximised values of average efficiency $\eta^{\odot}/\eta_C$ (red dashed line, left axis) and $R_{\eta}^{\Delta}$ (blue solid line, right axis) (b) as a function of total time of the cycle $\tau_u$ for a-QOE with a HO as the working substance. Parameter values are $\omega_2 = 2.0 \omega_1$, $\beta_h \omega_1 = 0.1$ and $\beta_c \omega_1 = 0.5$.}
	\label{fig:optimisationHO_eta}
\end{figure}
For the TLS case, the transition probability matrix for the work strokes can be written as:
\begin{align}
T_J(m,n)=\begin{pmatrix}
Q_J & 1-Q_J \\
1-Q_J& Q_J\\
\end{pmatrix},
\end{align}
where as before the index $J$ takes the values $I$ ($II$) on the left hand side and $f$ ($b$) on the right hand side for the compression (expansion) strokes. The values $Q_f$ and $Q_b$ are protocol dependent and are as described in Appendix \ref{app:A2}. 
For the thermal strokes described by Eq.~\eqref{eq:TLS dissipative master equation} for a TLS of energy gap $\Delta$, of duration $\tau$ in contact with a bath with temperature $\beta$, the transition probability in the energy basis can be written as Eq.~\eqref{eq:TLS finite time thermal transition probabilities}.With this, we can write the relevant transition matrices in Eq.~\eqref{eq:Tcycle Otto} for the TLS case as $T_{\beta_c}(n^\prime,l) = \bar{T}_{\beta_c}(n^\prime,l,(1-r_b)\tau_b,\omega_1)$ and $T_{\beta_h}(k,m) = \bar{T}_{\beta_h}(k,m,r_b\tau_b,\omega_2)$.

\begin{widetext}
\begin{align}
\label{eq:TLS finite time thermal transition probabilities}
\bar{T}_{\beta}(n,l,\tau,\Delta) =&\begin{pmatrix}
e^{-\gamma\tau} \left(1+e^{-\beta\Delta(e^{-\gamma\tau}-1)}\right) & e^{-\gamma\tau-\beta\Delta}(e^{-\gamma\tau}-1) \\
e^{-\gamma\tau-\beta\Delta}(e^{-\gamma\tau}-1)(e^{\beta\Delta}-1)& 1-e^{-\gamma\tau-\beta\Delta}(e^{-\gamma\tau}-1)\\
\end{pmatrix}
\end{align}
\end{widetext}
\section{\label{app:C} Optimization of Efficiency and its Fluctuations}
\begin{figure}
	\centering
	\subfloat{\begin{overpic}[width=0.9 \linewidth]{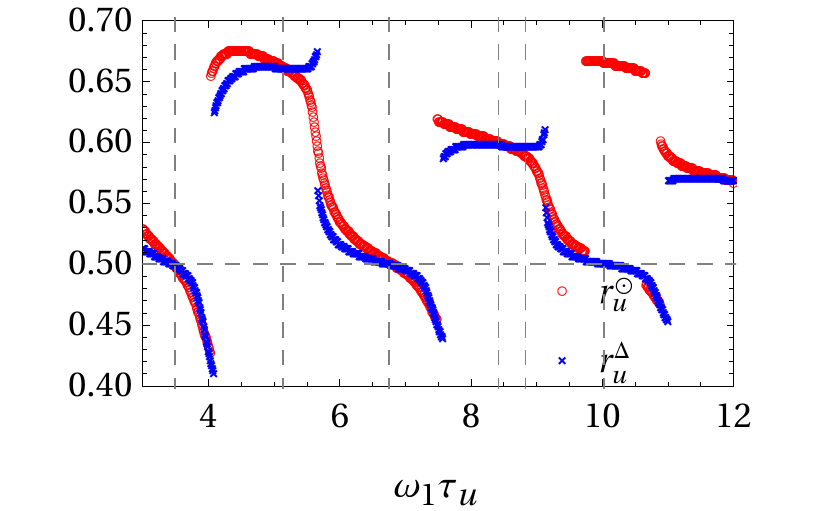}
	\put(20,50){\textbf{(a)}}
	\end{overpic}
	}\\ 
	\subfloat{\begin{overpic}[width=0.9 \linewidth]{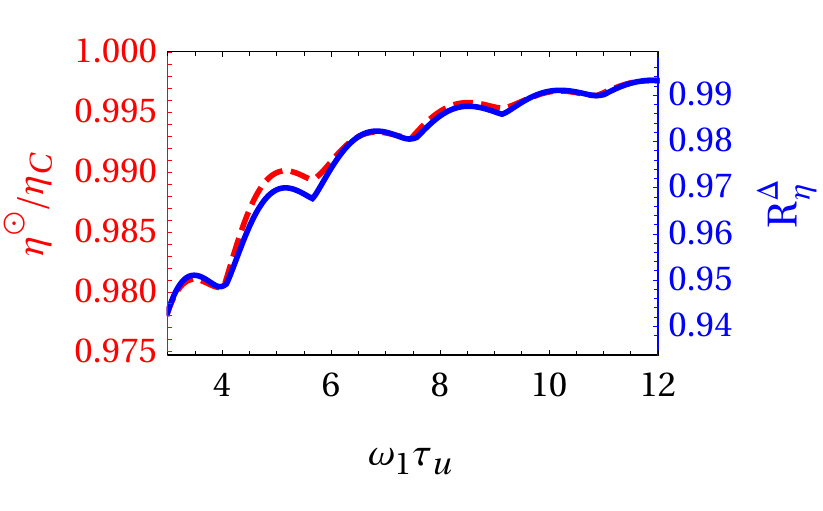}
	\put(20,50){\textbf{(b)}}
	\end{overpic}
	}
	\caption{(Color Online) Optimum values of the asymmetry parameters $r^{\odot}_{\eta}$ and $r^{\Delta}_{R_{\eta}}$ along with the corresponding maximised values of average efficiency $\eta^{\odot}/\eta_C$ (red dashed line, left axis) and $R_{\eta}^{\Delta}$ (blue solid line, right axis) (b) as a function of total time of the cycle $\tau_u$ for an a-QOE consists with a TLS as the working substance. Parameter values are $\omega_2 = 2.0 \omega_1$, $\delta=\omega_1$, $\beta_h \omega_1 = 0.1$, and $\beta_c \omega_1 = 0.5$.}
	\label{fig:optimisationTLS_eta}
\end{figure}
While we have focused on the optimization of work output and reliability in the main text, an important figure of merit of heat engines that is also of interest is the average efficiency defined as $\avg{\eta} = -\avg{w}/\avg{q_h}$. Moreover, an interesting parameter that characterizes the fluctuations of the efficiency is \cite{PhysRevE.103.L060103}
\begin{align}
    \eta^{(2)} = \frac{\sigma_w^2}{\avg{q_h^2}-\avg{q_h}^2}.
\end{align}
Just as we examined the average work output and reliability's behaviour as a function of the asymmetry parameter $r_u$, we now consider the optimal value of $r_u = r^{\odot}_{u}$ that maximises the efficiency $\eta$ as well as the values $r_u = r^{\Delta}_{u}$ that maximises the reliability of the efficiency defined as $R_\eta = \avg{\eta}/\sqrt{\eta^{(2)}}$. We restrict ourselves to perfect thermalization during the heat strokes. Figs.~\eqref{fig:optimisationHO_eta} (a) and \eqref{fig:optimisationTLS_eta} (a) clearly illustrate that like the optimal values for average work and reliability $r_u^{*},r_u^{\circ}$, the optimal $r_u$ values for efficiency and its reliability $r^{\odot},r^{\Delta}$ also show discontinuities as a function of $\tau_u$. Moreover, there are again specific values of $\tau_u$ where $r^{\odot}=r^{\Delta} $ i.e. the efficiency and its reliability are co-optimized. Finally, in Figs.~\eqref{fig:optimisationHO_eta} (b) and \eqref{fig:optimisationTLS_eta} (b) we see that in the limit of quasi-static work strokes with $\omega_1 \tau_u \gg 1$, as expected, the efficiency tends to the Carnot value $\eta_C$ and the reliability of efficiency saturates the bound $\avg{\eta}/\sqrt{\eta^{(2)}} \leq 1$ discovered in \cite{PhysRevE.103.L060103}.

\bibliography{bibliography}

\begin{thebibliography}{45}%
\makeatletter
\providecommand \@ifxundefined [1]{%
 \@ifx{#1\undefined}
}%
\providecommand \@ifnum [1]{%
 \ifnum #1\expandafter \@firstoftwo
 \else \expandafter \@secondoftwo
 \fi
}%
\providecommand \@ifx [1]{%
 \ifx #1\expandafter \@firstoftwo
 \else \expandafter \@secondoftwo
 \fi
}%
\providecommand \natexlab [1]{#1}%
\providecommand \enquote  [1]{``#1''}%
\providecommand \bibnamefont  [1]{#1}%
\providecommand \bibfnamefont [1]{#1}%
\providecommand \citenamefont [1]{#1}%
\providecommand \href@noop [0]{\@secondoftwo}%
\providecommand \href [0]{\begingroup \@sanitize@url \@href}%
\providecommand \@href[1]{\@@startlink{#1}\@@href}%
\providecommand \@@href[1]{\endgroup#1\@@endlink}%
\providecommand \@sanitize@url [0]{\catcode `\\12\catcode `\$12\catcode
  `\&12\catcode `\#12\catcode `\^12\catcode `\_12\catcode `\%12\relax}%
\providecommand \@@startlink[1]{}%
\providecommand \@@endlink[0]{}%
\providecommand \url  [0]{\begingroup\@sanitize@url \@url }%
\providecommand \@url [1]{\endgroup\@href {#1}{\urlprefix }}%
\providecommand \urlprefix  [0]{URL }%
\providecommand \Eprint [0]{\href }%
\providecommand \doibase [0]{http://dx.doi.org/}%
\providecommand \selectlanguage [0]{\@gobble}%
\providecommand \bibinfo  [0]{\@secondoftwo}%
\providecommand \bibfield  [0]{\@secondoftwo}%
\providecommand \translation [1]{[#1]}%
\providecommand \BibitemOpen [0]{}%
\providecommand \bibitemStop [0]{}%
\providecommand \bibitemNoStop [0]{.\EOS\space}%
\providecommand \EOS [0]{\spacefactor3000\relax}%
\providecommand \BibitemShut  [1]{\csname bibitem#1\endcsname}%
\let\auto@bib@innerbib\@empty
\bibitem [{\citenamefont {Scovil}\ and\ \citenamefont
  {Schulz-DuBois}(1959)}]{Scovil1959}%
  \BibitemOpen
  \bibfield  {author} {\bibinfo {author} {\bibfnamefont {H.~E.~D.}\
  \bibnamefont {Scovil}}\ and\ \bibinfo {author} {\bibfnamefont {E.~O.}\
  \bibnamefont {Schulz-DuBois}},\ }\href {\doibase 10.1103/PhysRevLett.2.262}
  {\bibfield  {journal} {\bibinfo  {journal} {Physical Review Letters}\
  }\textbf {\bibinfo {volume} {2}},\ \bibinfo {pages} {262} (\bibinfo {year}
  {1959})}\BibitemShut {NoStop}%
\bibitem [{\citenamefont {Alicki}(1979)}]{Alicki1979}%
  \BibitemOpen
  \bibfield  {author} {\bibinfo {author} {\bibfnamefont {R.}~\bibnamefont
  {Alicki}},\ }\href {\doibase 10.1088/0305-4470/12/5/007} {\bibfield
  {journal} {\bibinfo  {journal} {Journal of Physics A: Mathematical and
  General}\ }\textbf {\bibinfo {volume} {12}},\ \bibinfo {pages} {L103}
  (\bibinfo {year} {1979})}\BibitemShut {NoStop}%
\bibitem [{\citenamefont {Vinjanampathy}\ and\ \citenamefont
  {Anders}(2016)}]{Vinjanampathy2016}%
  \BibitemOpen
  \bibfield  {author} {\bibinfo {author} {\bibfnamefont {S.}~\bibnamefont
  {Vinjanampathy}}\ and\ \bibinfo {author} {\bibfnamefont {J.}~\bibnamefont
  {Anders}},\ }\href {\doibase 10.1080/00107514.2016.1201896} {\bibfield
  {journal} {\bibinfo  {journal} {Contemporary Physics}\ }\textbf {\bibinfo
  {volume} {57}},\ \bibinfo {pages} {545} (\bibinfo {year} {2016})}\BibitemShut
  {NoStop}%
\bibitem [{\citenamefont {Binder}\ \emph {et~al.}(2018)\citenamefont {Binder},
  \citenamefont {Correa}, \citenamefont {Gogolin}, \citenamefont {Anders},\
  and\ \citenamefont {Adesso}}]{Binder2018}%
  \BibitemOpen
  \bibinfo {editor} {\bibfnamefont {F.}~\bibnamefont {Binder}}, \bibinfo
  {editor} {\bibfnamefont {L.~A.}\ \bibnamefont {Correa}}, \bibinfo {editor}
  {\bibfnamefont {C.}~\bibnamefont {Gogolin}}, \bibinfo {editor} {\bibfnamefont
  {J.}~\bibnamefont {Anders}}, \ and\ \bibinfo {editor} {\bibfnamefont
  {G.}~\bibnamefont {Adesso}},\ eds.,\ \href {\doibase
  10.1007/978-3-319-99046-0} {\emph {\bibinfo {title} {{Thermodynamics in the
  Quantum Regime}}}},\ \bibinfo {series} {Fundamental Theories of Physics},
  Vol.\ \bibinfo {volume} {195}\ (\bibinfo  {publisher} {Springer International
  Publishing},\ \bibinfo {address} {Cham},\ \bibinfo {year} {2018})\BibitemShut
  {NoStop}%
\bibitem [{\citenamefont {Bhattacharjee}\ and\ \citenamefont
  {Dutta}(2021)}]{Bhattacharjee2021}%
  \BibitemOpen
  \bibfield  {author} {\bibinfo {author} {\bibfnamefont {S.}~\bibnamefont
  {Bhattacharjee}}\ and\ \bibinfo {author} {\bibfnamefont {A.}~\bibnamefont
  {Dutta}},\ }\href {\doibase 10.1140/epjb/s10051-021-00235-3} {\bibfield
  {journal} {\bibinfo  {journal} {The European Physical Journal B}\ }\textbf
  {\bibinfo {volume} {94}},\ \bibinfo {pages} {239} (\bibinfo {year}
  {2021})}\BibitemShut {NoStop}%
\bibitem [{\citenamefont {Mukherjee}\ and\ \citenamefont
  {Divakaran}(2021)}]{Mukherjee2021}%
  \BibitemOpen
  \bibfield  {author} {\bibinfo {author} {\bibfnamefont {V.}~\bibnamefont
  {Mukherjee}}\ and\ \bibinfo {author} {\bibfnamefont {U.}~\bibnamefont
  {Divakaran}},\ }\href {\doibase 10.1088/1361-648x/ac1b60} {\bibfield
  {journal} {\bibinfo  {journal} {Journal of Physics: Condensed Matter}\
  }\textbf {\bibinfo {volume} {33}},\ \bibinfo {pages} {454001} (\bibinfo
  {year} {2021})}\BibitemShut {NoStop}%
\bibitem [{\citenamefont {Myers}\ \emph {et~al.}(2022)\citenamefont {Myers},
  \citenamefont {Abah},\ and\ \citenamefont {Deffner}}]{Myers2022}%
  \BibitemOpen
  \bibfield  {author} {\bibinfo {author} {\bibfnamefont {N.~M.}\ \bibnamefont
  {Myers}}, \bibinfo {author} {\bibfnamefont {O.}~\bibnamefont {Abah}}, \ and\
  \bibinfo {author} {\bibfnamefont {S.}~\bibnamefont {Deffner}},\ }\href
  {http://arxiv.org/abs/2201.01740} {\ ,\ \bibinfo {pages} {1} (\bibinfo {year}
  {2022})}\BibitemShut {NoStop}%
\bibitem [{\citenamefont {Denzler}\ and\ \citenamefont
  {Lutz}(2020)}]{Denzler20}%
  \BibitemOpen
  \bibfield  {author} {\bibinfo {author} {\bibfnamefont {T.}~\bibnamefont
  {Denzler}}\ and\ \bibinfo {author} {\bibfnamefont {E.}~\bibnamefont {Lutz}},\
  }\href {\doibase 10.1103/PhysRevResearch.2.032062} {\bibfield  {journal}
  {\bibinfo  {journal} {Phys. Rev. Research}\ }\textbf {\bibinfo {volume}
  {2}},\ \bibinfo {pages} {032062} (\bibinfo {year} {2020})}\BibitemShut
  {NoStop}%
\bibitem [{\citenamefont {Saryal}\ \emph
  {et~al.}(2021{\natexlab{a}})\citenamefont {Saryal}, \citenamefont {Gerry},
  \citenamefont {Khait}, \citenamefont {Segal},\ and\ \citenamefont
  {Agarwalla}}]{Saryal21}%
  \BibitemOpen
  \bibfield  {author} {\bibinfo {author} {\bibfnamefont {S.}~\bibnamefont
  {Saryal}}, \bibinfo {author} {\bibfnamefont {M.}~\bibnamefont {Gerry}},
  \bibinfo {author} {\bibfnamefont {I.}~\bibnamefont {Khait}}, \bibinfo
  {author} {\bibfnamefont {D.}~\bibnamefont {Segal}}, \ and\ \bibinfo {author}
  {\bibfnamefont {B.~K.}\ \bibnamefont {Agarwalla}},\ }\href {\doibase
  10.1103/PhysRevLett.127.190603} {\bibfield  {journal} {\bibinfo  {journal}
  {Phys. Rev. Lett.}\ }\textbf {\bibinfo {volume} {127}},\ \bibinfo {pages}
  {190603} (\bibinfo {year} {2021}{\natexlab{a}})}\BibitemShut {NoStop}%
\bibitem [{\citenamefont {Saryal}\ and\ \citenamefont
  {Agarwalla}(2021{\natexlab{a}})}]{Saryal21a}%
  \BibitemOpen
  \bibfield  {author} {\bibinfo {author} {\bibfnamefont {S.}~\bibnamefont
  {Saryal}}\ and\ \bibinfo {author} {\bibfnamefont {B.~K.}\ \bibnamefont
  {Agarwalla}},\ }\href {\doibase 10.1103/PhysRevE.103.L060103} {\bibfield
  {journal} {\bibinfo  {journal} {Phys. Rev. E}\ }\textbf {\bibinfo {volume}
  {103}},\ \bibinfo {pages} {L060103} (\bibinfo {year}
  {2021}{\natexlab{a}})}\BibitemShut {NoStop}%
\bibitem [{\citenamefont {Barato}\ and\ \citenamefont
  {Seifert}(2015)}]{Barato15}%
  \BibitemOpen
  \bibfield  {author} {\bibinfo {author} {\bibfnamefont {A.~C.}\ \bibnamefont
  {Barato}}\ and\ \bibinfo {author} {\bibfnamefont {U.}~\bibnamefont
  {Seifert}},\ }\href {\doibase 10.1103/PhysRevLett.114.158101} {\bibfield
  {journal} {\bibinfo  {journal} {Phys. Rev. Lett.}\ }\textbf {\bibinfo
  {volume} {114}},\ \bibinfo {pages} {158101} (\bibinfo {year}
  {2015})}\BibitemShut {NoStop}%
\bibitem [{\citenamefont {Gingrich}\ \emph {et~al.}(2016)\citenamefont
  {Gingrich}, \citenamefont {Horowitz}, \citenamefont {Perunov},\ and\
  \citenamefont {England}}]{Gingrich16}%
  \BibitemOpen
  \bibfield  {author} {\bibinfo {author} {\bibfnamefont {T.~R.}\ \bibnamefont
  {Gingrich}}, \bibinfo {author} {\bibfnamefont {J.~M.}\ \bibnamefont
  {Horowitz}}, \bibinfo {author} {\bibfnamefont {N.}~\bibnamefont {Perunov}}, \
  and\ \bibinfo {author} {\bibfnamefont {J.~L.}\ \bibnamefont {England}},\
  }\href {\doibase 10.1103/PhysRevLett.116.120601} {\bibfield  {journal}
  {\bibinfo  {journal} {Phys. Rev. Lett.}\ }\textbf {\bibinfo {volume} {116}},\
  \bibinfo {pages} {120601} (\bibinfo {year} {2016})}\BibitemShut {NoStop}%
\bibitem [{\citenamefont {Timpanaro}\ \emph {et~al.}(2019)\citenamefont
  {Timpanaro}, \citenamefont {Guarnieri}, \citenamefont {Goold},\ and\
  \citenamefont {Landi}}]{Timpanaro19}%
  \BibitemOpen
  \bibfield  {author} {\bibinfo {author} {\bibfnamefont {A.~M.}\ \bibnamefont
  {Timpanaro}}, \bibinfo {author} {\bibfnamefont {G.}~\bibnamefont
  {Guarnieri}}, \bibinfo {author} {\bibfnamefont {J.}~\bibnamefont {Goold}}, \
  and\ \bibinfo {author} {\bibfnamefont {G.~T.}\ \bibnamefont {Landi}},\ }\href
  {\doibase 10.1103/PhysRevLett.123.090604} {\bibfield  {journal} {\bibinfo
  {journal} {Phys. Rev. Lett.}\ }\textbf {\bibinfo {volume} {123}},\ \bibinfo
  {pages} {090604} (\bibinfo {year} {2019})}\BibitemShut {NoStop}%
\bibitem [{\citenamefont {Pietzonka}\ and\ \citenamefont
  {Seifert}(2018)}]{Pietzonka18}%
  \BibitemOpen
  \bibfield  {author} {\bibinfo {author} {\bibfnamefont {P.}~\bibnamefont
  {Pietzonka}}\ and\ \bibinfo {author} {\bibfnamefont {U.}~\bibnamefont
  {Seifert}},\ }\href {\doibase 10.1103/PhysRevLett.120.190602} {\bibfield
  {journal} {\bibinfo  {journal} {Phys. Rev. Lett.}\ }\textbf {\bibinfo
  {volume} {120}},\ \bibinfo {pages} {190602} (\bibinfo {year}
  {2018})}\BibitemShut {NoStop}%
\bibitem [{\citenamefont {Saryal}\ \emph
  {et~al.}(2021{\natexlab{b}})\citenamefont {Saryal}, \citenamefont {Sadekar},\
  and\ \citenamefont {Agarwalla}}]{Saryal21c}%
  \BibitemOpen
  \bibfield  {author} {\bibinfo {author} {\bibfnamefont {S.}~\bibnamefont
  {Saryal}}, \bibinfo {author} {\bibfnamefont {O.}~\bibnamefont {Sadekar}}, \
  and\ \bibinfo {author} {\bibfnamefont {B.~K.}\ \bibnamefont {Agarwalla}},\
  }\href {\doibase 10.1103/PhysRevE.103.022141} {\bibfield  {journal} {\bibinfo
   {journal} {Phys. Rev. E}\ }\textbf {\bibinfo {volume} {103}},\ \bibinfo
  {pages} {022141} (\bibinfo {year} {2021}{\natexlab{b}})}\BibitemShut
  {NoStop}%
\bibitem [{\citenamefont {Feldmann}\ and\ \citenamefont
  {Kosloff}(2000)}]{Feldmann2000}%
  \BibitemOpen
  \bibfield  {author} {\bibinfo {author} {\bibfnamefont {T.}~\bibnamefont
  {Feldmann}}\ and\ \bibinfo {author} {\bibfnamefont {R.}~\bibnamefont
  {Kosloff}},\ }\href {\doibase 10.1103/PhysRevE.61.4774} {\bibfield  {journal}
  {\bibinfo  {journal} {Physical Review E}\ }\textbf {\bibinfo {volume} {61}},\
  \bibinfo {pages} {4774} (\bibinfo {year} {2000})}\BibitemShut {NoStop}%
\bibitem [{\citenamefont {Allahverdyan}\ \emph {et~al.}(2013)\citenamefont
  {Allahverdyan}, \citenamefont {Hovhannisyan}, \citenamefont {Melkikh},\ and\
  \citenamefont {Gevorkian}}]{Allahverdyan2013}%
  \BibitemOpen
  \bibfield  {author} {\bibinfo {author} {\bibfnamefont {A.~E.}\ \bibnamefont
  {Allahverdyan}}, \bibinfo {author} {\bibfnamefont {K.~V.}\ \bibnamefont
  {Hovhannisyan}}, \bibinfo {author} {\bibfnamefont {A.~V.}\ \bibnamefont
  {Melkikh}}, \ and\ \bibinfo {author} {\bibfnamefont {S.~G.}\ \bibnamefont
  {Gevorkian}},\ }\href {\doibase 10.1103/PhysRevLett.111.050601} {\bibfield
  {journal} {\bibinfo  {journal} {Physical Review Letters}\ }\textbf {\bibinfo
  {volume} {111}},\ \bibinfo {pages} {050601} (\bibinfo {year}
  {2013})}\BibitemShut {NoStop}%
\bibitem [{\citenamefont {Cavina}\ \emph {et~al.}(2017)\citenamefont {Cavina},
  \citenamefont {Mari},\ and\ \citenamefont {Giovannetti}}]{Cavina2017}%
  \BibitemOpen
  \bibfield  {author} {\bibinfo {author} {\bibfnamefont {V.}~\bibnamefont
  {Cavina}}, \bibinfo {author} {\bibfnamefont {A.}~\bibnamefont {Mari}}, \ and\
  \bibinfo {author} {\bibfnamefont {V.}~\bibnamefont {Giovannetti}},\ }\href
  {\doibase 10.1103/PhysRevLett.119.050601} {\bibfield  {journal} {\bibinfo
  {journal} {Physical Review Letters}\ }\textbf {\bibinfo {volume} {119}},\
  \bibinfo {pages} {050601} (\bibinfo {year} {2017})}\BibitemShut {NoStop}%
\bibitem [{\citenamefont {Abiuso}\ and\ \citenamefont
  {Perarnau-Llobet}(2020)}]{Abiuso2020}%
  \BibitemOpen
  \bibfield  {author} {\bibinfo {author} {\bibfnamefont {P.}~\bibnamefont
  {Abiuso}}\ and\ \bibinfo {author} {\bibfnamefont {M.}~\bibnamefont
  {Perarnau-Llobet}},\ }\href {\doibase 10.1103/PhysRevLett.124.110606}
  {\bibfield  {journal} {\bibinfo  {journal} {Physical Review Letters}\
  }\textbf {\bibinfo {volume} {124}},\ \bibinfo {pages} {110606} (\bibinfo
  {year} {2020})}\BibitemShut {NoStop}%
\bibitem [{\citenamefont {Dann}\ and\ \citenamefont
  {Kosloff}(2020)}]{Dann2020}%
  \BibitemOpen
  \bibfield  {author} {\bibinfo {author} {\bibfnamefont {R.}~\bibnamefont
  {Dann}}\ and\ \bibinfo {author} {\bibfnamefont {R.}~\bibnamefont {Kosloff}},\
  }\href {\doibase 10.1088/1367-2630/ab6876} {\bibfield  {journal} {\bibinfo
  {journal} {New Journal of Physics}\ }\textbf {\bibinfo {volume} {22}},\
  \bibinfo {pages} {013055} (\bibinfo {year} {2020})}\BibitemShut {NoStop}%
\bibitem [{\citenamefont {Rezek}\ and\ \citenamefont
  {Kosloff}(2006)}]{Rezek2006}%
  \BibitemOpen
  \bibfield  {author} {\bibinfo {author} {\bibfnamefont {Y.}~\bibnamefont
  {Rezek}}\ and\ \bibinfo {author} {\bibfnamefont {R.}~\bibnamefont
  {Kosloff}},\ }\href {\doibase 10.1088/1367-2630/8/5/083} {\bibfield
  {journal} {\bibinfo  {journal} {New Journal of Physics}\ }\textbf {\bibinfo
  {volume} {8}},\ \bibinfo {pages} {83} (\bibinfo {year} {2006})}\BibitemShut
  {NoStop}%
\bibitem [{\citenamefont {Quan}\ \emph {et~al.}(2007)\citenamefont {Quan},
  \citenamefont {Liu}, \citenamefont {Sun},\ and\ \citenamefont
  {Nori}}]{Quan2007}%
  \BibitemOpen
  \bibfield  {author} {\bibinfo {author} {\bibfnamefont {H.~T.}\ \bibnamefont
  {Quan}}, \bibinfo {author} {\bibfnamefont {Y.-x.}\ \bibnamefont {Liu}},
  \bibinfo {author} {\bibfnamefont {C.~P.}\ \bibnamefont {Sun}}, \ and\
  \bibinfo {author} {\bibfnamefont {F.}~\bibnamefont {Nori}},\ }\href {\doibase
  10.1103/PhysRevE.76.031105} {\bibfield  {journal} {\bibinfo  {journal}
  {Physical Review E}\ }\textbf {\bibinfo {volume} {76}},\ \bibinfo {pages}
  {031105} (\bibinfo {year} {2007})}\BibitemShut {NoStop}%
\bibitem [{\citenamefont {Abah}\ \emph {et~al.}(2012)\citenamefont {Abah},
  \citenamefont {Ro{\ss}nagel}, \citenamefont {Jacob}, \citenamefont {Deffner},
  \citenamefont {Schmidt-Kaler}, \citenamefont {Singer},\ and\ \citenamefont
  {Lutz}}]{Abah2012}%
  \BibitemOpen
  \bibfield  {author} {\bibinfo {author} {\bibfnamefont {O.}~\bibnamefont
  {Abah}}, \bibinfo {author} {\bibfnamefont {J.}~\bibnamefont {Ro{\ss}nagel}},
  \bibinfo {author} {\bibfnamefont {G.}~\bibnamefont {Jacob}}, \bibinfo
  {author} {\bibfnamefont {S.}~\bibnamefont {Deffner}}, \bibinfo {author}
  {\bibfnamefont {F.}~\bibnamefont {Schmidt-Kaler}}, \bibinfo {author}
  {\bibfnamefont {K.}~\bibnamefont {Singer}}, \ and\ \bibinfo {author}
  {\bibfnamefont {E.}~\bibnamefont {Lutz}},\ }\href {\doibase
  10.1103/PhysRevLett.109.203006} {\bibfield  {journal} {\bibinfo  {journal}
  {Physical Review Letters}\ }\textbf {\bibinfo {volume} {109}},\ \bibinfo
  {pages} {203006} (\bibinfo {year} {2012})}\BibitemShut {NoStop}%
\bibitem [{\citenamefont {Karimi}\ and\ \citenamefont
  {Pekola}(2016)}]{Karimi2016}%
  \BibitemOpen
  \bibfield  {author} {\bibinfo {author} {\bibfnamefont {B.}~\bibnamefont
  {Karimi}}\ and\ \bibinfo {author} {\bibfnamefont {J.~P.}\ \bibnamefont
  {Pekola}},\ }\href {\doibase 10.1103/PhysRevB.94.184503} {\bibfield
  {journal} {\bibinfo  {journal} {Physical Review B}\ }\textbf {\bibinfo
  {volume} {94}},\ \bibinfo {pages} {184503} (\bibinfo {year}
  {2016})}\BibitemShut {NoStop}%
\bibitem [{\citenamefont {Kosloff}\ and\ \citenamefont
  {Rezek}(2017)}]{Kosloff2017}%
  \BibitemOpen
  \bibfield  {author} {\bibinfo {author} {\bibfnamefont {R.}~\bibnamefont
  {Kosloff}}\ and\ \bibinfo {author} {\bibfnamefont {Y.}~\bibnamefont
  {Rezek}},\ }\href {\doibase 10.3390/e19040136} {\bibfield  {journal}
  {\bibinfo  {journal} {Entropy}\ }\textbf {\bibinfo {volume} {19}},\ \bibinfo
  {pages} {136} (\bibinfo {year} {2017})}\BibitemShut {NoStop}%
\bibitem [{\citenamefont {Chen}\ \emph {et~al.}(2019)\citenamefont {Chen},
  \citenamefont {Sun},\ and\ \citenamefont {Dong}}]{Chen2019}%
  \BibitemOpen
  \bibfield  {author} {\bibinfo {author} {\bibfnamefont {J.~F.}\ \bibnamefont
  {Chen}}, \bibinfo {author} {\bibfnamefont {C.~P.}\ \bibnamefont {Sun}}, \
  and\ \bibinfo {author} {\bibfnamefont {H.}~\bibnamefont {Dong}},\ }\href
  {\doibase 10.1103/PhysRevE.100.032144} {\bibfield  {journal} {\bibinfo
  {journal} {Physical Review E}\ }\textbf {\bibinfo {volume} {100}},\ \bibinfo
  {pages} {032144} (\bibinfo {year} {2019})}\BibitemShut {NoStop}%
\bibitem [{\citenamefont {Das}\ and\ \citenamefont
  {Mukherjee}(2020)}]{Das2020}%
  \BibitemOpen
  \bibfield  {author} {\bibinfo {author} {\bibfnamefont {A.}~\bibnamefont
  {Das}}\ and\ \bibinfo {author} {\bibfnamefont {V.}~\bibnamefont
  {Mukherjee}},\ }\href {\doibase 10.1103/PhysRevResearch.2.033083} {\bibfield
  {journal} {\bibinfo  {journal} {Physical Review Research}\ }\textbf {\bibinfo
  {volume} {2}},\ \bibinfo {pages} {033083} (\bibinfo {year}
  {2020})}\BibitemShut {NoStop}%
\bibitem [{\citenamefont {Insinga}(2020)}]{Insinga2020}%
  \BibitemOpen
  \bibfield  {author} {\bibinfo {author} {\bibfnamefont {A.~R.}\ \bibnamefont
  {Insinga}},\ }\href {\doibase 10.3390/E22091060} {\bibfield  {journal}
  {\bibinfo  {journal} {Entropy}\ }\textbf {\bibinfo {volume} {22}} (\bibinfo
  {year} {2020}),\ 10.3390/E22091060}\BibitemShut {NoStop}%
\bibitem [{\citenamefont {Singh}\ and\ \citenamefont {Abah}(2020)}]{Singh2020}%
  \BibitemOpen
  \bibfield  {author} {\bibinfo {author} {\bibfnamefont {S.}~\bibnamefont
  {Singh}}\ and\ \bibinfo {author} {\bibfnamefont {O.}~\bibnamefont {Abah}},\
  }\href {http://arxiv.org/abs/2008.05002} {\bibfield  {journal} {\bibinfo
  {journal} {arXiv:2008.05002}\ } (\bibinfo {year} {2020})}\BibitemShut
  {NoStop}%
\bibitem [{\citenamefont {Zhang}(2020)}]{Zhang2020}%
  \BibitemOpen
  \bibfield  {author} {\bibinfo {author} {\bibfnamefont {Y.}~\bibnamefont
  {Zhang}},\ }\href {\doibase 10.1016/j.physa.2020.125083} {\bibfield
  {journal} {\bibinfo  {journal} {Physica A: Statistical Mechanics and its
  Applications}\ }\textbf {\bibinfo {volume} {559}},\ \bibinfo {pages} {125083}
  (\bibinfo {year} {2020})}\BibitemShut {NoStop}%
\bibitem [{\citenamefont {Cavina}\ \emph {et~al.}(2021)\citenamefont {Cavina},
  \citenamefont {Erdman}, \citenamefont {Abiuso}, \citenamefont {Tolomeo},\
  and\ \citenamefont {Giovannetti}}]{Cavina2021}%
  \BibitemOpen
  \bibfield  {author} {\bibinfo {author} {\bibfnamefont {V.}~\bibnamefont
  {Cavina}}, \bibinfo {author} {\bibfnamefont {P.~A.}\ \bibnamefont {Erdman}},
  \bibinfo {author} {\bibfnamefont {P.}~\bibnamefont {Abiuso}}, \bibinfo
  {author} {\bibfnamefont {L.}~\bibnamefont {Tolomeo}}, \ and\ \bibinfo
  {author} {\bibfnamefont {V.}~\bibnamefont {Giovannetti}},\ }\href {\doibase
  10.1103/PhysRevA.104.032226} {\bibfield  {journal} {\bibinfo  {journal}
  {Physical Review A}\ }\textbf {\bibinfo {volume} {104}},\ \bibinfo {pages}
  {32226} (\bibinfo {year} {2021})}\BibitemShut {NoStop}%
\bibitem [{\citenamefont {{Terr{\'{e}}n Alonso}}\ \emph
  {et~al.}(2022)\citenamefont {{Terr{\'{e}}n Alonso}}, \citenamefont {Abiuso},
  \citenamefont {Perarnau-Llobet},\ and\ \citenamefont
  {Arrachea}}]{TerrenAlonso2022}%
  \BibitemOpen
  \bibfield  {author} {\bibinfo {author} {\bibfnamefont {P.}~\bibnamefont
  {{Terr{\'{e}}n Alonso}}}, \bibinfo {author} {\bibfnamefont {P.}~\bibnamefont
  {Abiuso}}, \bibinfo {author} {\bibfnamefont {M.}~\bibnamefont
  {Perarnau-Llobet}}, \ and\ \bibinfo {author} {\bibfnamefont {L.}~\bibnamefont
  {Arrachea}},\ }\href {\doibase 10.1103/prxquantum.3.010326} {\bibfield
  {journal} {\bibinfo  {journal} {PRX Quantum}\ }\textbf {\bibinfo {volume}
  {3}},\ \bibinfo {pages} {1} (\bibinfo {year} {2022})}\BibitemShut {NoStop}%
\bibitem [{\citenamefont {Ashida}\ and\ \citenamefont
  {Sagawa}(2021)}]{Ashida2021}%
  \BibitemOpen
  \bibfield  {author} {\bibinfo {author} {\bibfnamefont {Y.}~\bibnamefont
  {Ashida}}\ and\ \bibinfo {author} {\bibfnamefont {T.}~\bibnamefont
  {Sagawa}},\ }\href {\doibase 10.1038/s42005-021-00553-z} {\bibfield
  {journal} {\bibinfo  {journal} {Communications Physics}\ }\textbf {\bibinfo
  {volume} {4}},\ \bibinfo {pages} {45} (\bibinfo {year} {2021})}\BibitemShut
  {NoStop}%
\bibitem [{\citenamefont {Erdman}\ and\ \citenamefont
  {No{\'{e}}}(2021)}]{Erdman2021}%
  \BibitemOpen
  \bibfield  {author} {\bibinfo {author} {\bibfnamefont {P.~A.}\ \bibnamefont
  {Erdman}}\ and\ \bibinfo {author} {\bibfnamefont {F.}~\bibnamefont
  {No{\'{e}}}},\ }\href {\doibase 10.1038/s41534-021-00512-0} {\bibfield
  {journal} {\bibinfo  {journal} {npj Quantum Information}\ }\textbf {\bibinfo
  {volume} {8}},\ \bibinfo {pages} {1} (\bibinfo {year} {2021})}\BibitemShut
  {NoStop}%
\bibitem [{\citenamefont {Khait}\ \emph {et~al.}(2021)\citenamefont {Khait},
  \citenamefont {Carrasquilla},\ and\ \citenamefont {Segal}}]{Khait2021}%
  \BibitemOpen
  \bibfield  {author} {\bibinfo {author} {\bibfnamefont {I.}~\bibnamefont
  {Khait}}, \bibinfo {author} {\bibfnamefont {J.}~\bibnamefont {Carrasquilla}},
  \ and\ \bibinfo {author} {\bibfnamefont {D.}~\bibnamefont {Segal}},\ }\href
  {http://arxiv.org/abs/2108.12441} {\bibfield  {journal} {\bibinfo  {journal}
  {arXiv:2108.12441}\ } (\bibinfo {year} {2021})}\BibitemShut {NoStop}%
\bibitem [{\citenamefont {Zheng}\ \emph {et~al.}(2016)\citenamefont {Zheng},
  \citenamefont {H\"anggi},\ and\ \citenamefont
  {Poletti}}]{PhysRevE.94.012137}%
  \BibitemOpen
  \bibfield  {author} {\bibinfo {author} {\bibfnamefont {Y.}~\bibnamefont
  {Zheng}}, \bibinfo {author} {\bibfnamefont {P.}~\bibnamefont {H\"anggi}}, \
  and\ \bibinfo {author} {\bibfnamefont {D.}~\bibnamefont {Poletti}},\ }\href
  {\doibase 10.1103/PhysRevE.94.012137} {\bibfield  {journal} {\bibinfo
  {journal} {Phys. Rev. E}\ }\textbf {\bibinfo {volume} {94}},\ \bibinfo
  {pages} {012137} (\bibinfo {year} {2016})}\BibitemShut {NoStop}%
\bibitem [{\citenamefont {Gorini}\ \emph {et~al.}(1976)\citenamefont {Gorini},
  \citenamefont {Kossakowski},\ and\ \citenamefont {Sudarshan}}]{Gorini1976}%
  \BibitemOpen
  \bibfield  {author} {\bibinfo {author} {\bibfnamefont {V.}~\bibnamefont
  {Gorini}}, \bibinfo {author} {\bibfnamefont {A.}~\bibnamefont {Kossakowski}},
  \ and\ \bibinfo {author} {\bibfnamefont {E.~C.~G.}\ \bibnamefont
  {Sudarshan}},\ }\href {\doibase 10.1063/1.522979} {\bibfield  {journal}
  {\bibinfo  {journal} {Journal of Mathematical Physics}\ }\textbf {\bibinfo
  {volume} {17}},\ \bibinfo {pages} {821} (\bibinfo {year} {1976})}\BibitemShut
  {NoStop}%
\bibitem [{\citenamefont {Lindblad}(1976)}]{Lindblad1976}%
  \BibitemOpen
  \bibfield  {author} {\bibinfo {author} {\bibfnamefont {G.}~\bibnamefont
  {Lindblad}},\ }\href {\doibase 10.1007/BF01608499} {\bibfield  {journal}
  {\bibinfo  {journal} {Communications in Mathematical Physics}\ }\textbf
  {\bibinfo {volume} {48}},\ \bibinfo {pages} {119} (\bibinfo {year}
  {1976})}\BibitemShut {NoStop}%
\bibitem [{\citenamefont {Feldmann}\ and\ \citenamefont
  {Kosloff}(2004)}]{Feldmann2004}%
  \BibitemOpen
  \bibfield  {author} {\bibinfo {author} {\bibfnamefont {T.}~\bibnamefont
  {Feldmann}}\ and\ \bibinfo {author} {\bibfnamefont {R.}~\bibnamefont
  {Kosloff}},\ }\href {\doibase 10.1103/PhysRevE.70.046110} {\bibfield
  {journal} {\bibinfo  {journal} {Physical Review E}\ }\textbf {\bibinfo
  {volume} {70}},\ \bibinfo {pages} {046110} (\bibinfo {year}
  {2004})}\BibitemShut {NoStop}%
\bibitem [{\citenamefont {Deffner}\ and\ \citenamefont
  {Lutz}(2008)}]{PhysRevE.77.021128}%
  \BibitemOpen
  \bibfield  {author} {\bibinfo {author} {\bibfnamefont {S.}~\bibnamefont
  {Deffner}}\ and\ \bibinfo {author} {\bibfnamefont {E.}~\bibnamefont {Lutz}},\
  }\href {\doibase 10.1103/PhysRevE.77.021128} {\bibfield  {journal} {\bibinfo
  {journal} {Phys. Rev. E}\ }\textbf {\bibinfo {volume} {77}},\ \bibinfo
  {pages} {021128} (\bibinfo {year} {2008})}\BibitemShut {NoStop}%
\bibitem [{\citenamefont {Ding}\ \emph {et~al.}(2018)\citenamefont {Ding},
  \citenamefont {Yi}, \citenamefont {Kim},\ and\ \citenamefont
  {Talkner}}]{PhysRevE.98.042122}%
  \BibitemOpen
  \bibfield  {author} {\bibinfo {author} {\bibfnamefont {X.}~\bibnamefont
  {Ding}}, \bibinfo {author} {\bibfnamefont {J.}~\bibnamefont {Yi}}, \bibinfo
  {author} {\bibfnamefont {Y.~W.}\ \bibnamefont {Kim}}, \ and\ \bibinfo
  {author} {\bibfnamefont {P.}~\bibnamefont {Talkner}},\ }\href {\doibase
  10.1103/PhysRevE.98.042122} {\bibfield  {journal} {\bibinfo  {journal} {Phys.
  Rev. E}\ }\textbf {\bibinfo {volume} {98}},\ \bibinfo {pages} {042122}
  (\bibinfo {year} {2018})}\BibitemShut {NoStop}%
\bibitem [{\citenamefont {Dann}\ \emph {et~al.}(2020)\citenamefont {Dann},
  \citenamefont {Kosloff},\ and\ \citenamefont {Salamon}}]{e22111255}%
  \BibitemOpen
  \bibfield  {author} {\bibinfo {author} {\bibfnamefont {R.}~\bibnamefont
  {Dann}}, \bibinfo {author} {\bibfnamefont {R.}~\bibnamefont {Kosloff}}, \
  and\ \bibinfo {author} {\bibfnamefont {P.}~\bibnamefont {Salamon}},\ }\href
  {https://www.mdpi.com/1099-4300/22/11/1255} {\bibfield  {journal} {\bibinfo
  {journal} {Entropy}\ }\textbf {\bibinfo {volume} {22}},\ \bibinfo {pages}
  {1255} (\bibinfo {year} {2020})}\BibitemShut {NoStop}%
\bibitem [{\citenamefont {Plastina}\ \emph {et~al.}(2014)\citenamefont
  {Plastina}, \citenamefont {Alecce}, \citenamefont {Apollaro}, \citenamefont
  {Falcone}, \citenamefont {Francica}, \citenamefont {Galve}, \citenamefont
  {Lo~Gullo},\ and\ \citenamefont {Zambrini}}]{Plastina2014}%
  \BibitemOpen
  \bibfield  {author} {\bibinfo {author} {\bibfnamefont {F.}~\bibnamefont
  {Plastina}}, \bibinfo {author} {\bibfnamefont {A.}~\bibnamefont {Alecce}},
  \bibinfo {author} {\bibfnamefont {T.~J.~G.}\ \bibnamefont {Apollaro}},
  \bibinfo {author} {\bibfnamefont {G.}~\bibnamefont {Falcone}}, \bibinfo
  {author} {\bibfnamefont {G.}~\bibnamefont {Francica}}, \bibinfo {author}
  {\bibfnamefont {F.}~\bibnamefont {Galve}}, \bibinfo {author} {\bibfnamefont
  {N.}~\bibnamefont {Lo~Gullo}}, \ and\ \bibinfo {author} {\bibfnamefont
  {R.}~\bibnamefont {Zambrini}},\ }\href {\doibase
  10.1103/PhysRevLett.113.260601} {\bibfield  {journal} {\bibinfo  {journal}
  {Phys. Rev. Lett.}\ }\textbf {\bibinfo {volume} {113}},\ \bibinfo {pages}
  {260601} (\bibinfo {year} {2014})}\BibitemShut {NoStop}%
\bibitem [{\citenamefont {Alecce}\ \emph {et~al.}(2015)\citenamefont {Alecce},
  \citenamefont {Galve}, \citenamefont {Gullo}, \citenamefont {Dell'Anna},
  \citenamefont {Plastina},\ and\ \citenamefont {Zambrini}}]{Alecce2015}%
  \BibitemOpen
  \bibfield  {author} {\bibinfo {author} {\bibfnamefont {A.}~\bibnamefont
  {Alecce}}, \bibinfo {author} {\bibfnamefont {F.}~\bibnamefont {Galve}},
  \bibinfo {author} {\bibfnamefont {N.~L.}\ \bibnamefont {Gullo}}, \bibinfo
  {author} {\bibfnamefont {L.}~\bibnamefont {Dell'Anna}}, \bibinfo {author}
  {\bibfnamefont {F.}~\bibnamefont {Plastina}}, \ and\ \bibinfo {author}
  {\bibfnamefont {R.}~\bibnamefont {Zambrini}},\ }\href {\doibase
  10.1088/1367-2630/17/7/075007} {\bibfield  {journal} {\bibinfo  {journal}
  {New Journal of Physics}\ }\textbf {\bibinfo {volume} {17}},\ \bibinfo
  {pages} {075007} (\bibinfo {year} {2015})}\BibitemShut {NoStop}%
\bibitem [{\citenamefont {Saryal}\ and\ \citenamefont
  {Agarwalla}(2021{\natexlab{b}})}]{PhysRevE.103.L060103}%
  \BibitemOpen
  \bibfield  {author} {\bibinfo {author} {\bibfnamefont {S.}~\bibnamefont
  {Saryal}}\ and\ \bibinfo {author} {\bibfnamefont {B.~K.}\ \bibnamefont
  {Agarwalla}},\ }\href {\doibase 10.1103/PhysRevE.103.L060103} {\bibfield
  {journal} {\bibinfo  {journal} {Phys. Rev. E}\ }\textbf {\bibinfo {volume}
  {103}},\ \bibinfo {pages} {L060103} (\bibinfo {year}
  {2021}{\natexlab{b}})}\BibitemShut {NoStop}%
\end{thebibliography}%

\end{document}